\documentclass[lettersize,journal]{IEEEtran}
\usepackage{amsmath,amsfonts}
\usepackage{algorithmic}
\usepackage{algorithm}
\usepackage{array}
\usepackage{subfig}
\usepackage{textcomp}
\usepackage{stfloats}
\usepackage{url}
\usepackage{verbatim}
\usepackage{graphicx}
\usepackage{cite}
\usepackage{bm} 
\usepackage{amssymb} 

\begin{document}

\title{\fontsize{22pt}{30pt}\selectfont Beamforming Design for Wideband Near-Field Communications With Reconfigurable Refractive Surfaces}

\author{\vspace{0.2cm}Zicheng Lin, Shuhao Zeng,~\IEEEmembership{Member,~IEEE}, Aryan Kaushik,~\IEEEmembership{Member,~IEEE}, \\and Hongliang Zhang,~\IEEEmembership{Member,~IEEE} \vspace{-0.8cm}
\thanks{Part of this work has been accepted by IEEE WCNC 2025~\cite{wcnc1}.}
\thanks{Zicheng Lin is with the School of Electronic and Computer Engineering, Peking University, Beijing 1008711, China (e-mail: zicheng.lin@stu.pku.edu.com).}
\thanks{Shuhao Zeng is with the School of Electronic and Computer Engineering, Peking University Shenzhen Graduate School, Shenzhen 518055, China (e-mail: shuhao.zeng@pku.edu.cn).}
\thanks{Aryan Kaushik is with the Department of Computing and Mathematics, Manchester Metropolitan University, UK (e-mail: a.kaushik@mmu.ac.uk).}
\thanks{Hongliang Zhang, is with the School of Electronics, Peking University, Beijing 100871, China (e-mail: hongliang.zhang@ pku.edu.cn).}
}



\maketitle

\begin{abstract}
To meet the growing demand for high data rates, cellular systems are expected to evolve towards higher carrier frequencies and larger antenna arrays, but conventional phased arrays face challenges in supporting such a prospection due to their excessive power consumption induced by numerous phase shifters required. Reconfigurable Refractive Surface (RRS) is an energy efficient solution to address this issue without relying on phase shifters. However, the increased radiation aperture size extends the range of the Fresnel region, leading the users to lie in the near-field zone. Moreover, given the wideband communications in higher frequency bands, we cannot ignore the frequency selectivity of the RRS. These two effects collectively exacerbate the beam-split issue, where different frequency components fail to converge on the user simultaneously, and finally result in a degradation of the data rate. In this paper, we investigate an RRS-based wideband near-field multi-user communication system. Unlike most existing studies on wideband communications, which consider the beam-split effect only with the near-field condition, we study the beam-split effect under the influence of both the near-field condition and the frequency selectivity of the RRS. To mitigate the beam-split effect, we propose a Delayed-RRS structure, based on which a beamforming scheme is proposed to optimize the user's data rate. Through theoretical analysis and simulation results, we analyze the influence of the RRS's frequency selectivity, demonstrate the effectiveness of the proposed beamforming scheme, and reveal the importance of jointly considering the near-field condition and the frequency selectivity of RRS.
\end{abstract}


\section{Introduction}
As effective approaches to achieve higher data rates, larger antenna arrays and higher communication frequency bands have gained increasing attention due to various emerging 6G applications. However, traditional phased arrays face significant challenges in supporting such large-scale deployments and high-frequency communications for its excessive power consumption induced by numerous phase shifters.
Recently, metasurface-based antennas, such as the reconfigurable refractive surfaces (RRSs) mentioned in~\cite{zengshuhao1}, provide an energy efficient alternative to phased arrays~\cite{aryan1,aryan2}. The RRS is a surface consisting of numerous sub-wavelength elements. Each element can apply phase shifts to the signals refracted through it and thereby behave like a phase shifter. There are several low-power diodes on each element and we can modify the phase shift by controlling the biased voltage applied to the  diodes to realize the desired beamforming at an ultra-low power consumption~\cite{zhangshuhang,liang2}. 

Nevertheless, the enlarged radiation aperture of the RRS and the higher communication frequency will lead to an expanded near-field range, making the users more likely to fall in the near field of the RRS. Moreover, a broader bandwidth allowed by the higher frequency band also results in considerable frequency selectivity of RRS~\cite{resonant1,practical2}, which is manifested by different transmission coefficeints at different subcarrier frequencies. These two features have a great impact on the beam split problem that occurs in traditional wideband communication systems. To be specific, the beam split problem refers to the phenomenon that the beams on different subcarriers would point to different directions. On the one hand, in the near field, the beams on different subcarriers would focus on different locations rather than point to different directions. On the other hand, the extra phase errors between different subcarriers introduced by the frequency selectivity of the RRS will result in a loss of focus for the beams on some subcarriers. As a result, the beam split issue become more complex and severe.

Existing works related to the wideband metasurface-assisted communications focus on mitigating the beam split effect while considering either the near-field ~\cite{wide3,wide4,wide5,double} or the frequency selectivity of the metasurfaces~\cite{cairismodel,cai2,rismodel2}. In~\cite{wide3}, the authors investigated the distinctions of the beam split effect in the near-field region of metasurfaces, which leads to the degradation of data rates. To address this issue, the authors introduced a sub-array architecture and proposed a deep learning-based optimization framework to optimize the achievable communication rate. 
In~\cite{wide4}, the authors researched an extremely large-scale metasurface-based antenna array. With considering the near-field spherical wave model, the authors applied the delay adjustable metasurfaces to alleviate the near-field beam split effect, and an optimization framework was proposed to maximize the system sum rate. In~\cite{wide5}, the authors analyzed the upper bound of a metasurface-aided near-field communication system and proposed a quadratic optimization framework for properly setting the phase shifts of the metasurface.

In addition, a few works have noticed the frequency selectivity of metasurfaces which also leads to the beam split problem~\cite{cairismodel,cai2,rismodel2}. In~\cite{cairismodel}, based on the equivalent circuit theory, the authors constructed a model to estimate the frequency selective response of metasurfaces, and analyzed the performance loss caused by ignoring the frequency selectivity. In~\cite{cai2}, considering a frequency selective metasurface-assisted multi-user multi-band communication system, the authors proposed a more optimization-friendly metasurface model and improved the system's communication rate by solving the power minimization and sum-rate maximization problems.  In~\cite{rismodel2}, the authors considered a wideband metasurface-empowered wireless system with multiple users, in which the element response of metasurfaces is modeled according to the Lorentzian form. The authors maximized the achievable sum rate through optimizing the parameters dictating the Lorentzian response of each metasurface's element, and quantified the performance loss for assuming the frequency-flat response of metasurfaces. 
However, these works have not realized that the effects of the near-field channel conditions and the metasurface's frequency selectivity on the beam split problem are simultaneously present and mutually influential. Ignoring either factor would result in significant performance degradation.

In this paper, we consider an RRS-based wideband near-field multi-user communication system. When the signals propagate from the frequency selective RRS to the users in the near-field region, both the near-field channel and the frequency selectivity of the RRS would introduce different phase shifts and amplitude variations to the signals over different subcarrier frequencies, which poses significant challenges for wideband beamforming.
To compensate for the frequency selectivity, we introduce time-delay units in RRS elements. To imporve the data rate of the system, we propose an new beamforming scheme which jointly optimizes the performance loss caused by the near-field channel condition and the frequency selectivity of the RRS. 
Specifically, the main contributions of this paper are summarized as follows:
\begin{itemize}
    \item We consider a wideband multi-user downlink network where an RRS is used as the transmit antennas in the BS. The frequency selective transmission coefficient of RRS is modeled, based on which, we reveal the new characteristic of the beam split effect under the near-field channel condition and the frequency selectivity of RRS. Time delay units are introduced in the RRS elements to mitigate the beam split effect.

    \item Based on the known channel imformation, we analyze the data rate of each user. An optimization problem is formulated to maximize the minimum data rate of all users by jointly optimizing the digital beamformer, the phase shifts of RRS elements and the time delay of the delay units. Then we propose an effective alternate optimization algorithm to solve the formulated problem.
    
    \item Based on the proposed beamforming scheme, We analyze the performance of the RRS-based communication system. Simulation results verify the effectiveness of our proposed scheme. In addition, the impacts of the RRS scale, the distance between the feeds and the RRS, the number of the feeds and the number of quantization bits of RRS elements are demonstrated numerically.
    
\end{itemize}

The rest of the paper is organized as follows. In Section II, the system models are introduced. In Section III, we introduce the beam split effect in the near field of a frequency selective RRS and the delay-RRS structure. In Section IV, a optimization problem is formulated to maximize the minimum rate of all users, which is decomposed into three subproblems. In section V, an efficient algorithm is proposed to solve the formulated problems. In section VI, we analyze the performance of the proposed algorithm and the base station's structure. Simulation results are provided in Section VII. Finally, conclusion remarks are drawn in Section VIII.

\section{System Model}

In this section, we first introduce a RRS-based MIMO system where the base station (BS) adopts an RRS as the transmit antennas. Next, the models of the RRS and the channel in wideband near-field communication scenario are constructed. Further, the RRS-based hybrid beamforming architecture is introduced and finally a mathematical model of the signals received by the users is proposed. 

\subsection{Scenario Description}

\begin{figure}[ht]
\vspace{-0.1cm}
\centerline{\hspace{0.1cm}\includegraphics[width=8.5cm]{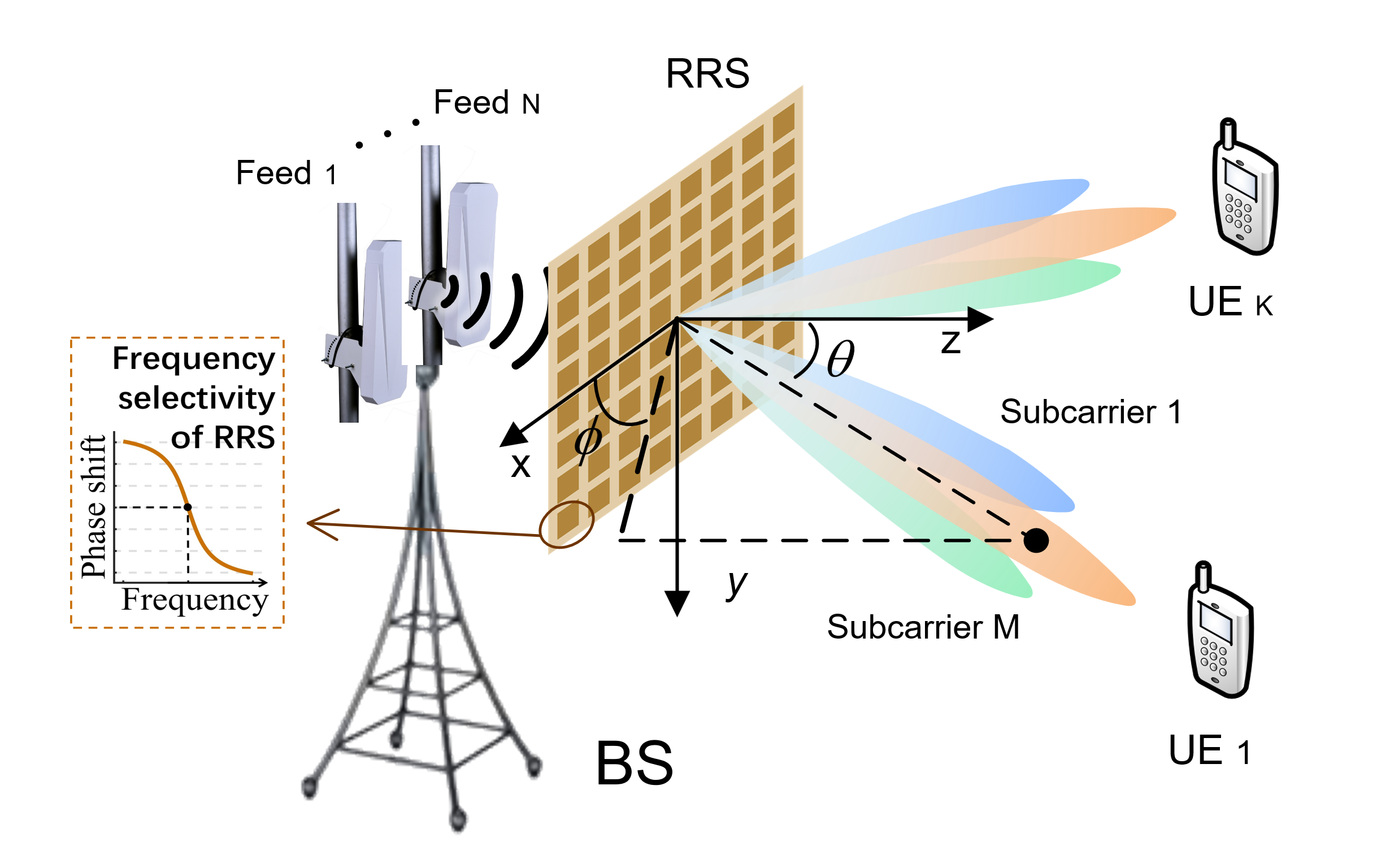}}
\vspace{0.1cm}
\caption{System model of a wideband downlink multi-user network, where an RRS is used as the BS antenna.}
\vspace{0.1cm}
\label{scenario}
\end{figure}

As shown in Fig.~\ref{scenario}, we consider an RRS-enabled wideband downlink communication system, where one BS serves $K$ users. In the BS, an RRS illuminated by $N$ feeds is adopted as the transmit antenna~\cite{zengshuhao1}. In order to point the transmitted wave of the feeds toward the RRS, each feed is assumed to be directional. Each user is equipped with an omni-directional antenna. 

To avoid the interference between symbols caused by the wideband communication, we adopt the OFDM technique to divide the bandwidth $W$ into $M$ subcarriers. Thus we can assume that the signal on each subcarrier is a narrow-band signal, i.e., the channel response can be considered invariant within the bandwidth of a subcarrier. The center frequency is denoted as $f_c$, and the frequency of the $m^{th}$ subcarrier is $f_{m}=f_{c}+\frac{W(2m-1-M)}{2M}$. 


\subsection{Reconfigurable Refractive Surfaces}
The RRS is a type of reconfigurable intelligent surface that can manipulate the propagation of incident electromagnetic waves to improve the system performance~\cite{liang1}.
It consists of $N_a$ rows and $N_b$ columns of sub-wavelength elements, and the total number of the RRS elements is represented as $N_R=N_a\times N_b$.
Different from the traditional reflective metasurface element as referred in~\cite{ris1}, the RRS element does not contain the copper backplane so that most of the incident signals can penetrate the elements~\cite{rrs1,rrs2}. 
Each RRS element contains several diodes, which can be either positive-intrinsic-negative (PIN)~\cite{pin} or varactor diodes~\cite{varactor}. By changing the biased voltages applied to the diodes, the transmission coefficient of each RRS element can change accordingly. 

Note that each metasurface element is equivalent to a resonant circuit~\cite{resonant1}. Therefore, in the wideband communication scenario, the frequency selectivity of the RRS cannot be ignored, namely, the change of each element’s transmission coefficient with frequency. However, since the bandwidth of each subcarrier is relatively narrow, the transmission coefficient within each subcarrier can be regarded as a consistent~\cite{antenna}. Consequently, we use $\psi_{m,i}$ to represent the transmission coefficient of the $i^{th}$ RRS element over the $m^{th}$ subcarrier. Because of the equivalent circuit characteristic, the transmission coefficients over different subcarriers are coupled with each other, i.e., when the biased voltage of the RRS element becomes different, its transmission coefficient over all subcarriers, including transmission phase shift and transmission amplitude, would change accordingly.

Let $\theta_i^c$ denote the phase shift brought by the $i^{th}$ RRS element at the center frequency. Since $\theta_i^c$ is a function of the bias voltage, 
we can use $\theta_i^c$ to characterize the influence of the bias voltage~\cite{practical2}. 
Therefore, the transmission coefficient of the $i^{th}$ RRS element over the $m^{th}$ subcarrier $\psi_{m,i}$ can be modeled as\vspace{-0.15cm}
\begin{equation} 
    \psi_{m,i}=\psi(\theta_i^c,f_m),
    \label{psim}
\end{equation} 


\subsection{Channel Model}
Due to the its high energy efficiency, a large RRS can be deeployed as the transmit antenna to provide significant beamforming gain. However, the increased radiation aperture size extends the range of the Fresnel region, leading the users to lie in the near-field zone~\cite{zhouzhou}. To characterize the near-field channel, we adopt the non-uniform spherical wave (NUSW) model, which considers the non-uniform power and the spherical wavefront~\cite{unidis,zhouzhou}. 
We use $\mathbf{H}_{m}=[\mathbf{h}_{m,1},...,\mathbf{h}_{m,K}]^T\in \mathbb{C}^{K\times N_R}$ to denote the channel matrix between RRS and $K$ users over the $m^{th}$ subcarrier, where $\mathbf{h}_{m,k}$ denotes the channel between the $k^{th}$ user and the RRS. We assume line of sight (LoS) propagations between the RRS and the users where the small scale fading is ignored for ease of analysis. Therefore, the channel between $i^{th}$ RRS element and $k^{th}$ user can be given by~\cite{rrs1,tangwankai}
\begin{equation}
    [\mathbf{h}_{m,k}]_i = \sqrt{\frac{G_rF_uA_r}{4\pi r^2_{i,k}}} e^{-j2\pi\frac{f_m}{c}r_{i,k}},\label{channel1}
\end{equation}
where $G_r$ denotes the transmit gain pattern of the $i^{th}$ RRS element in the direction of the $k^{th}$ user, $F_u$ represents the normalized radiation pattern of the $k^{th}$ user's antenna in the direction of the $i^{th}$ RRS element, $r_{i,k}$ denotes the distance between the $i^{th}$ RRS element and the $k^{th}$ user, and $A_r$ denotes the aperture of the user's antenna. Since the user's antenna is omni-directional, it may be assumed that $F_u = 1$. Further, the transmit gain pattern of the RRS element $G_r$ can be given by~\cite{radi_pattern}:
\begin{equation}
    G_r=G(\alpha_r)\begin{cases} 2(\alpha_r+1)\cos^{\alpha_r}\theta, \theta\in[0,\frac{\pi}{2}],\\ 0, \text{otherwise,}\end{cases}
    \label{radi_pattern}
\end{equation}
where $2(\alpha_r+1)$ denotes the gain of the RRS element and $\theta$ denotes the angle between the orientation of the RRS and the direction from the $i^{th}$ RRS element towards the $k^{th}$ user. \footnote{It assumes isotropy in directions parallel to the RRS.}

\subsection{Hybrid Beamforming}
To direct the transmitted beams towards different users, a hybrid beamforming architecture is adopted, where the transmitted data is first encoded with a digital beamforming matrix, and then fed to the RRS for analog beamforming by cofiguring its phase shifts. 

1) Digital Beamforming at the BS: 
Let $\mathbf{s}_m = \left[s_{m,1},\cdots,s_{m,K}\right]^T\in \mathbb{C}^{K\times 1}$ be the transmit symbols for $K$ different users over the $m^{th}$ subcarrier. The symbol vector $\mathbf{s}_m$ is first precoded by the digital beamformer $\mathbf{G}_m,\ m\in\{1,\cdots,M\}  $. Each $\mathbf{G}_m=[\mathbf{g}_{m,1},...,\mathbf{g}_{m,K}]\in\mathbb{C}^{N\times K}$ denotes the digital beamformer on the $m^{th}$ subcarrier, where $\mathbf{g}_{m,k}\in \mathbb{C}^{N\times 1}$ is used to precode the symbol $s_{m,k}$ into $N$ encoded signals. 
In order to ensure that the transmitted $K$ symbols on each subcarrier can be restored, the number of the RF chains should be larger than the number of symbols, i.e., $N\geq K$ should be satisfied. Then the encoded signals are up-converted to the $m^{th}$ carrier frequency and delivered to the $N$ feeds respectively. Let $\mathbf{s}^{\prime}_m\in \mathbb{C}^{N\times 1}$ denote the transmitted signals from the feeds, which can be expressed as
\begin{equation}
    \mathbf{s}^{\prime}_m = \mathbf{G}_m\mathbf{s}_m.
    \label{sig1}
\end{equation}

\begin{figure*}[t]
    \vspace*{-0cm}
    \centering
        \subfloat[Far-field, frequency independent RRS]{\includegraphics[height=3.9cm]{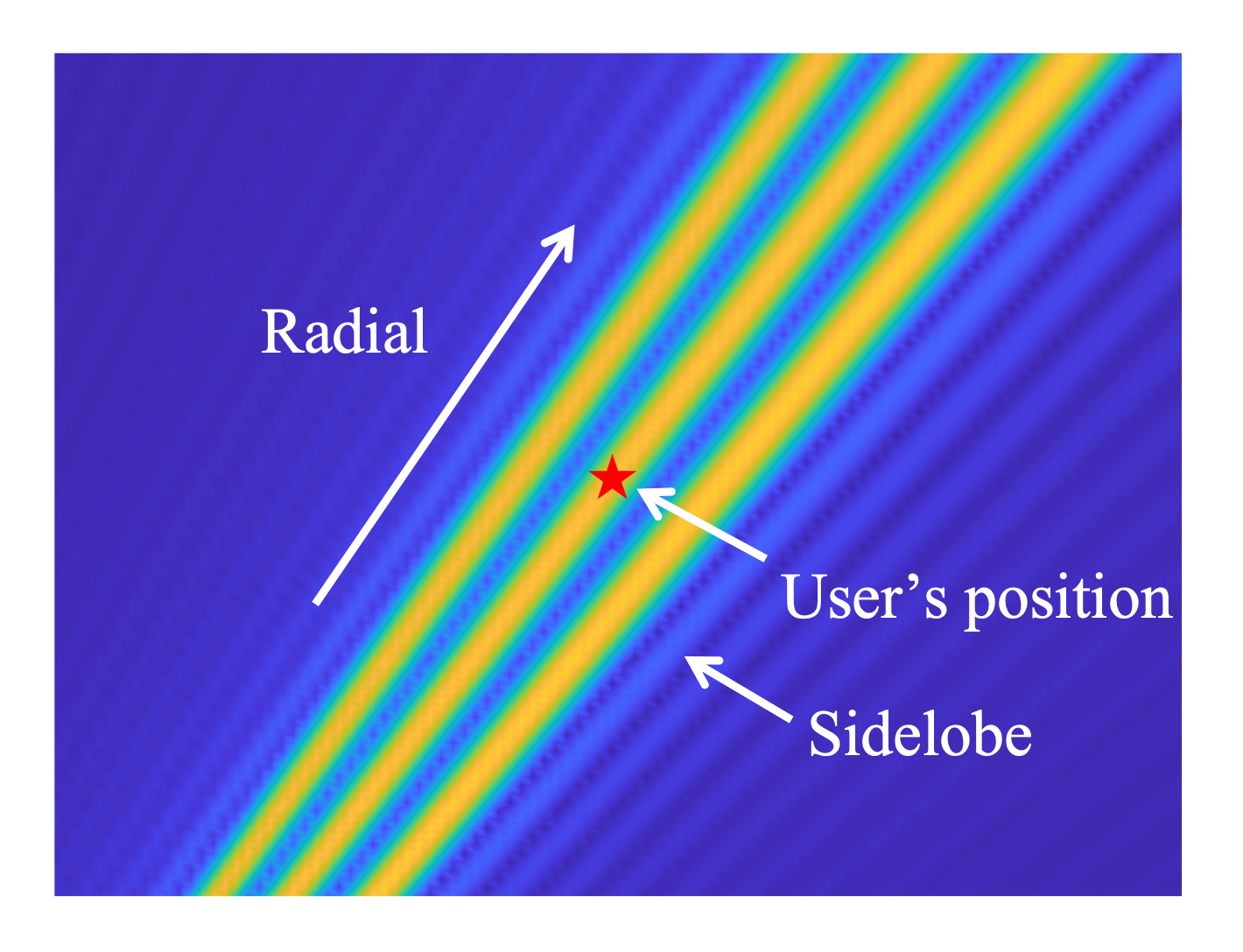}}
        \hspace{0.1cm}
        \subfloat[Near-field, frequency independent RRS]{\includegraphics[height=3.9cm]{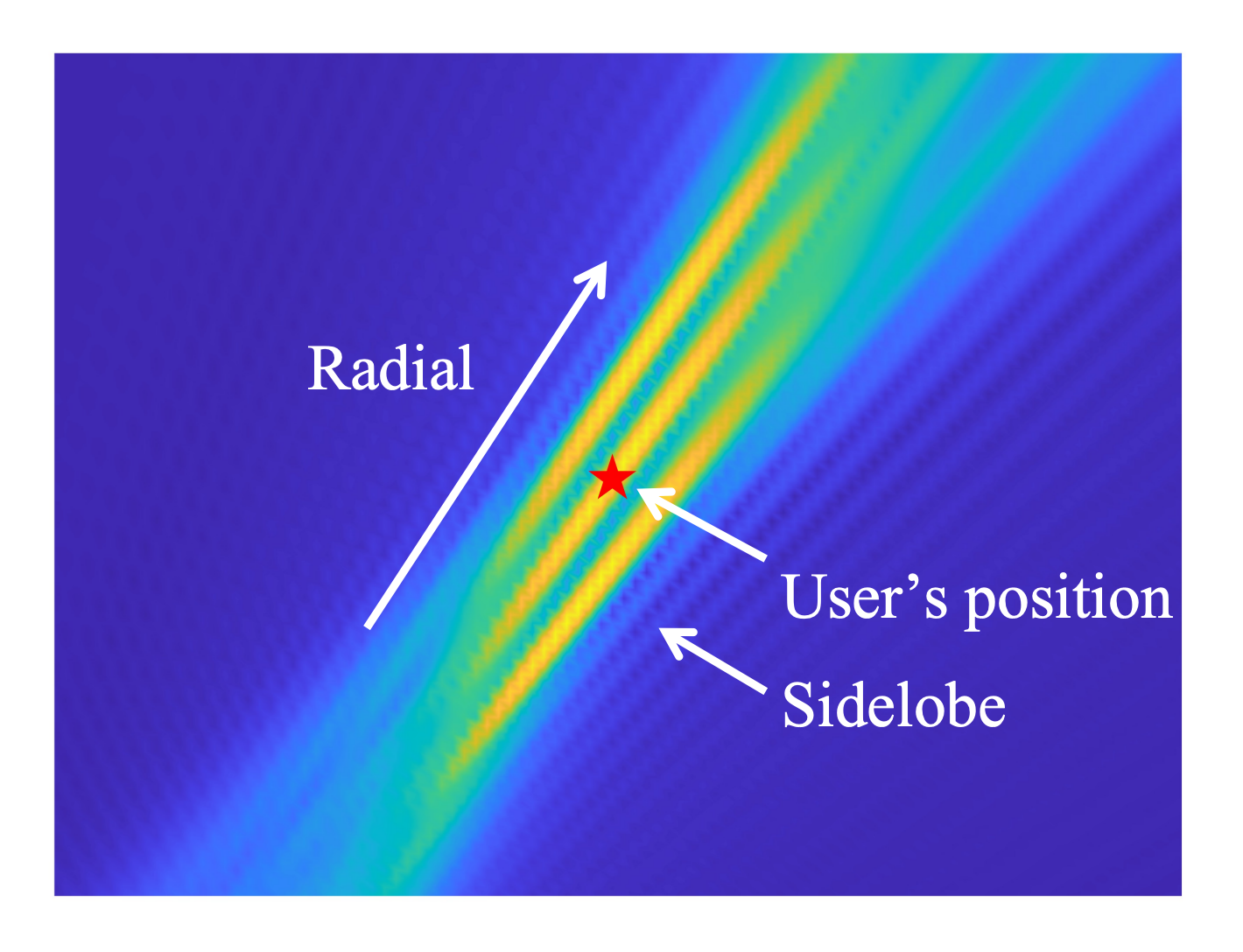}}
        \hspace{0.1cm}
        \subfloat[Near-field, frequency selective RRS \quad\quad\quad]{\includegraphics[height=3.9cm]{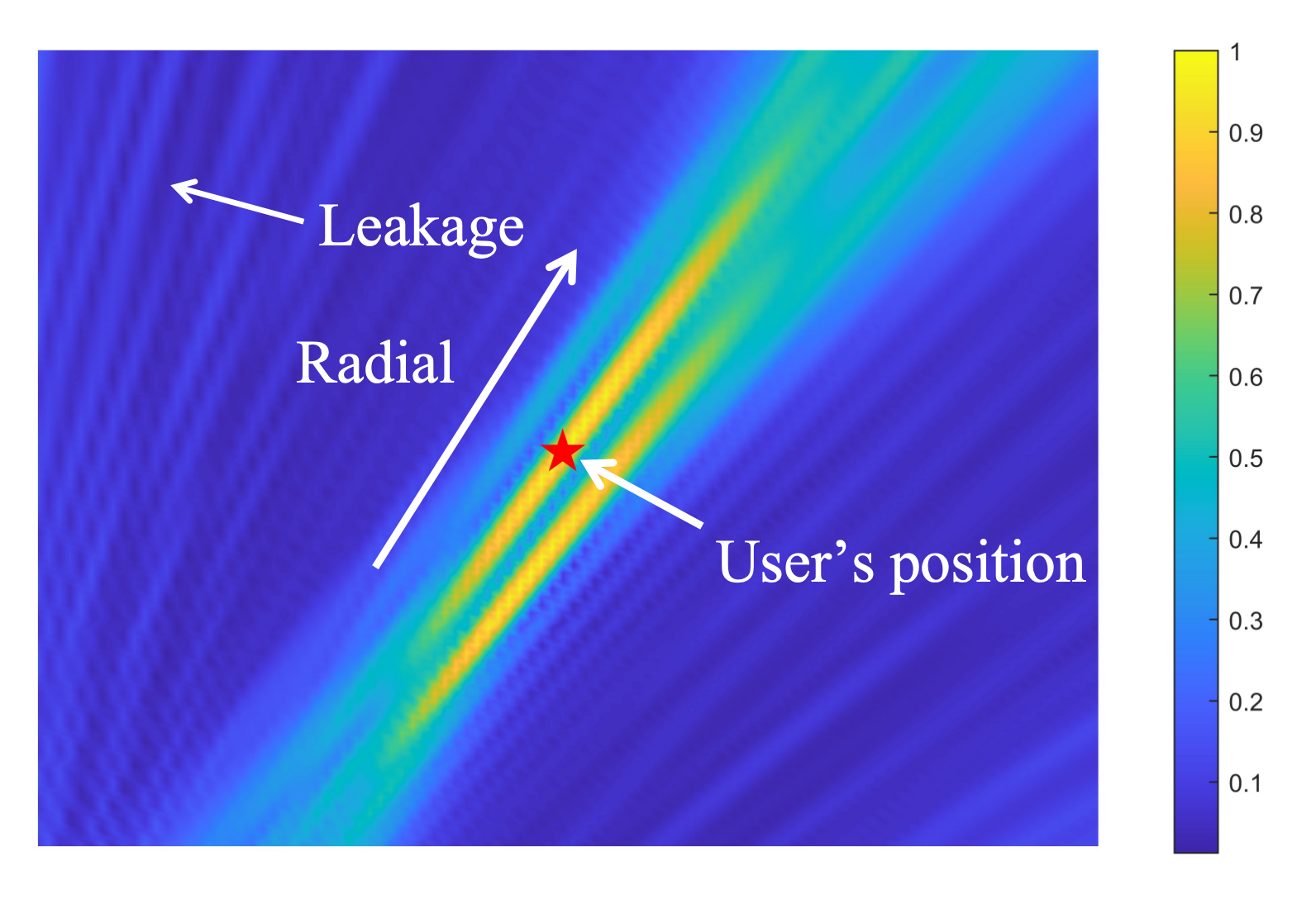}} 
    \vspace{-0.0cm}
    \caption{Beam Split Phenomenons}\label{beamsp}
    \vspace*{-0.5cm}
\end{figure*}

2) Analog Beamforming at the RRS: 
When the signals transmitted by the feeds impinge upon the RRS elements, the signals will be refracted. The refracted signal over the $m^{th}$ subcarrier  $\mathbf{x}_m\in \mathbb{C}^{N_R\times 1}$ can be expressed as
\begin{equation}
    \mathbf{x}_m = \boldsymbol{\Psi}_m\mathbf{B}_m\mathbf{s}^{\prime}_m.
    \label{sig2}
\end{equation}
Here, $\mathbf{B}_m\in \mathbb{C}^{N_R\times N}$ denotes the effect of wireless environments on the transmitted signal $\mathbf{s}^{\prime}_m$ when it propagates from the feeds to the RRS, 
the matrix $\boldsymbol{\Psi}_m$ denotes the transmission coefficient of the whole RRS array over the $m^{th}$ subcarrier, which is given by stacking the transmission coefficeints $\psi_{m,i}$ of all RRS elements into a diagonal matrix, i.e., $\bm{\Psi}_m=diag\{\psi_{m,1},...,\psi_{m,N_R}\}\in \mathbb{C}^{N_R\times N_R}$.

For compactness and allowing the RRS to capture most of the energy radiated by the feeds, we assume that the feeds are in the near field of the RRS. Thus we adopt the NUSW model to characterize the radio propagation characteristic from the $j^{th}$ feed to the $i^{th}$ RRS element over the $m^{th}$ subcarrier, i.e.,~\cite{rrs1,tangwankai}
\begin{equation}
    [\mathbf{B}_m]_{i,j} = \sqrt{\frac{G_tF_rS_0}{4\pi d^2_{i,j}}} e^{-j2\pi\frac{f_m}{c}d_{i,j}},
\end{equation}
where $G_t$ denotes the transmit gain pattern of the $j^{th}$ feed in the direction of the $i^{th}$ RRS element, $F_r$ represents the normalized radiation pattern of the $i^{th}$ RRS element in the direction of the $j^{th}$ feed, $S_0$ represents the area of each RRS element, and $d_{i,j}$ denotes the distance between the $i^{th}$ RRS element and the $j^{th}$ feed. 

Similarly, the transmit gain pattern of the feed antenna $G_t$ can be given by $G_t = G(\alpha_t)$ as~\eqref{radi_pattern}, where $\alpha_t$ represents the directivity of the feed antenna, and the normalized radiation pattern of the RRS element $F_r$ can be given by~\cite{tangwankai} 
\begin{equation}
    \left.F_r=\begin{cases} \cos^{\alpha_r}\theta, \theta\in[0,\frac{\pi}{2}],\\ 0, \text{otherwise,}\end{cases}\right.
    \label{fr}
\end{equation}
where $\theta$ denotes the angle between the orientation of the RRS and the direction from the $i^{th}$ RRS element towards the $j^{th}$ feed antenna.

\subsection{Signal Model} 
The refracted signals $\mathbf{x}_m$ propagate through the channels $\mathbf{h}_{m,k}$ from the RRS to the users, where the signal received at the $k^{th}$ user over the $m^{th}$ subcarrier can be given as follows according to \eqref{sig1}, \eqref{sig2}
\begin{equation}
\begin{aligned}
    {y}_{m,k}&=\mathbf{h}_{m,k}^T\mathbf{x}_m+n_{m,k} \\
    &=\mathbf{h}_{m,k}^T\boldsymbol{\Psi}_m\mathbf{B}_m\mathbf{G}_m\mathbf{s}_m+n_{m,k},
    \label{recsig}
\end{aligned}
\end{equation}
where $\mathbf{s}_m$ denotes the transmitted symbols over the $m^{th}$ subcarrier, ${n}_{m,k}\sim \mathcal{CN}(0, \sigma_{m,k}^2)$ denotes the additive white Gaussian noise with zero mean and variance of $\sigma_{m,k}^2$. Further, let $s_{m,k}$ denotes the $k^{th}$ element of $\mathbf{s}_m$, i.e., the symbol transmitted for the $k^{th}$ user, 
then the received signal in \eqref{recsig} can be rewritten as
\begin{equation}
\begin{aligned}
	{y}_{m,k} = \ \  & \mathbf{h}_{m,k}^T\boldsymbol{\Psi}_m\mathbf{B}_m\mathbf{g}_{m,k}s_{m,k}+  \\
	&\sum_{i\neq k}\mathbf{h}_{m,k}^T\boldsymbol{\Psi}_m\mathbf{B}_m\mathbf{g}_{m,i}{s}_{m,i}+{n}_{m,k},
\end{aligned}
\end{equation}
where the first term on the right side of the equation denotes the signal required by the $k^{th}$ user, and the second term on the right side denotes the inter-user interference. Assuming that the information symbols $\mathbf{s}_m$ satisfy $\mathbb{E}[\mathbf{s}_m\mathbf{s}_m^H]=\mathbf{I}_K$, the signal-to-interference-plus-noise ratio (SINR) for the $k^{th}$ user over the $m^{th}$  subcarrier is given by 
\begin{equation}
    \gamma_{m,k} = \frac{\left|\mathbf{h}_{m,k}^T\boldsymbol{\Psi}_m\mathbf{B}_m\mathbf{g}_{m,k}\right|^2}{\sigma_{m,k}^2+\sum^K_{i=1,i\neq k}\left|\mathbf{h}_{m,k}^T\boldsymbol{\Psi}_m\mathbf{B}_m\mathbf{g}_{m,i}\right|^2}. \label{sinr}
\end{equation}    
Then the data-rate of user $k$ can be given by
\begin{equation}
    R_{k} = \sum_{m=1}^{M}\log_2 \left( 1+\gamma_{m,k}\right).
\end{equation}

\section{Beam Split Problem and Its Compensation}
Similar to~\cite{steering,neardif}, in the RRS-based wideband communication scenario, the response of wireless channels varies with frequency, whose variation differs from that of the transmission coefficients of RRS elements. This mismatch prevents the RRS element from simultaneously compensating for the phase shifts of signals at each subcarrier frequency during propagation.  As a result, the beam split phenomenon still exists, and the signal power at different frequencies cannot all be directed toward the user.
In the following, we first demonstrate such beam split phenomenon in the near field of a frequency selective RRS in Section~\ref{aaa}. Then a delayed-RRS structure is proposed to mitigate the beam split effect in Section~\ref{bbb} .


\subsection{Beam Split Phenomenon}\label{aaa}
In this section, we use array gain as a metric to demonstrate the beam split phenomenon with considering the frequency selectivity of the RRS~\cite{steering2}. The array gain represents the gain in the signal strength resulting from the coherent superposition of signals transmitted from multiple antennas of an array comparing with transmitting signals with a single antenna. Based on the system model, the array gain at the user's location over the $m^{th}$ subcarrier can be given by 
\begin{equation}
    \rho(f_m,\mathbf{r}) = \left|\bm{\psi}_m\cdot \mathbf{a}(f_m,\mathbf{r})\right|.
    \label{rrsarraygain}
\end{equation}
where $f_m$ denotes the frequency of the $m^{th}$ subcarrier and $\mathbf{r}$ denotes the user's location relative to the RRS. According to~\eqref{psim}, let $\bm{\psi}_m = \left[\psi_{m,1},...,\psi_{m,N_R}\right]^T $ denote the transmission coefficeints of the RRS, i.e. the phase shifts and amplitude gains applied to the refracted signal by the RRS. The $\mathbf{a}(f_m,\mathbf{r})$ denotes the steering vector at the $m^{th}$ subcarrier frequency, which characterizes the difference of the phase and amplitude between the signals transmitted from different RRS elements to the user.
For a user in the near field of the RRS, the steering vector $\mathbf{a}(f_m,\mathbf{r})$ at frequency $f_m$ can be given by
\begin{equation}
    \mathbf{a}(f_m,\mathbf{r})= \frac{\mathbf{h}(f_m,\mathbf{r})}{\sqrt{N_R}a_{m}},
\end{equation} 
where $\mathbf{r}$ denotes the user's location relative to the RRS, $\mathbf{h}(f_m,\mathbf{r})$ represents the channel vector from the RRS to the user according to~\eqref{channel1}, $N_R$ denotes the number of the RRS elements, $\frac{1}{\sqrt{N_R}}$ represents the normalization factor required to evenly distribute the transmit power among $N_R$ array elements, and $a_m$ represents the normalization factor used to eliminate the impact of the distance between the user and the RRS, which is equal to the path loss from the center of RRS to the user. 

To maxmize the array gain at the user's location over the center subcarrier, we set the phase shift of the $i^{th}$ element at the center frequency as \footnote{For simplicity, we derive the beamforming vector based on the assumption that the amplitude of the RRS element is little affected by the phase configuration, i.e.$|\psi(\theta^c_1,f_c)|\approx \cdots \approx |\psi(\theta^c_N,f_c)|$.} 
\begin{equation}
    \theta^c_i = 2\pi \frac{f_c}{c}d_i,\ \forall i,
\end{equation}
where $d_i$ denotes the distance between the $i^{th}$ element of the RRS and the user.
Therefore, the phases of the signals at the center frequency received by the user from different RRS units are the same.

To illustrate the beam split phenomenon in the near-field of an RRS,  we provide a numerical case study in Fig.~\ref{beamsp} (c), where
we consider a uniform linear array consisting of frequency selective RRS elements and plot the array gain distribution of three subcarriers in the 2-D plane. 
In addition, as shown in Fig.~\ref{beamsp} (a) and Fig.~\ref{beamsp} (b), in order to highlight the uniqueness of the beam split phenomenon in a RRS-based wideband near-field system, we also present the beam split phenomenon in the far field and near field of a frequency independent phased array for comparison. Similarly, the configuration of each phased array is optimized to maximize the array gain at the center frequency. 


As shown in Fig.~\ref{beamsp} (a), in the far field, the beams on different subcarriers would point towards different directions. Because the same beamforming vector is adopted on different subcarriers by the frequency independent phased array, the corresponding steering vectors at different subcarriers are the same too.  However, the steering vector is related to the frequency and the direction in the far-field approximation. When the frequency changes, the corresponding direction will change accordingly.

In contrast, as shown in Fig.~\ref{beamsp} (b), different beams would focus on different locations in the near field. This is because, the more accurate spherical wavefront is considered in near-field scenarios, which means the steering vector is not only related to the direction but also related to the distance. Thus only near the target point can the array gain be large enough, which is reflected in the figure as the beam is focused on locations.

In Fig.~\ref{beamsp} (c), the beam split phenomenon differs from Fig.~\ref{beamsp} (a) and Fig.~\ref{beamsp} (b) in that, the beams lose focus on the edge subcarriers only in the RRS scenario. This is because the frequency selectivity of the RRS introduces extra phase errors on different subcarriers, which disrupt the phase relationships on the edge subcarriers among the elements in the beamforming vector, which results in that the upper beam almost disappears and the more severe signal leakage happens in the Fig.~\ref{beamsp} (c). 

Due to the beam split effects, the user can only access to the signals close to the center frequency, which is even more serious in the RRS-enabled communication scenario. This effect eventually has a negative impact on the communication rates of the RRS-enabled wideband near-field communication systems. 

\subsection{Hardware Compensation with Delayed-RRS}\label{bbb}
To mitigate the above issue, we propose a new architecture of the RRS, called Delayed-RRS. We introduce the time-delay unit (TD) in the RRS element~\cite{delayrrs}, which can compensate for the extra phase shift caused by the frequency selectivity. 
\begin{figure}[h]
    \vspace{-0.35cm}
    \centerline{\includegraphics[width=7.2cm]{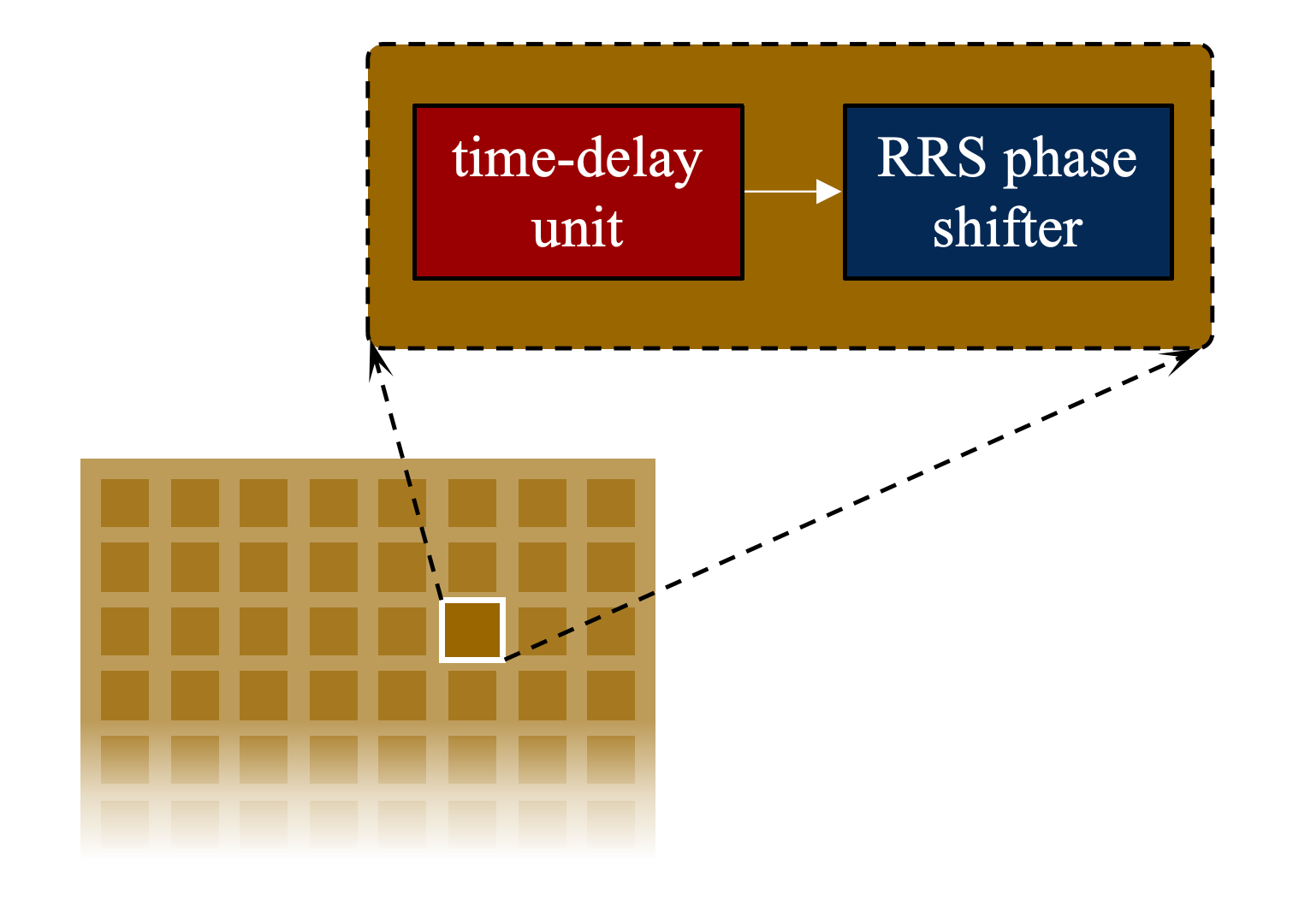}}
    \vspace{-0.3cm}
    \caption{The equivalent structure of the RRS element.}
    \vspace{-0.2cm}
    \label{RRSTDD}
\end{figure}

As shown in Fig.~\ref{RRSTDD}, each element of delayed-RRS is equivalent to the combination of a time-delay unit and a traditional RRS element. Then the transmission coefficient of the $i^{th}$ RRS element can be rewriten as 

\begin{equation}
    \psi_{m,i}=e^{-j2\pi f_m\tau_i}\psi(\theta_i^c,f_m),
\end{equation}
where $\tau_i$ denotes the $i^{th}$ element's delay and $\psi(\cdot)$ given in~\eqref{psim} denotes the transmission coefficient of traditional RRS elements.

Existing RRS-based beamforming schemes neglect the frequency selectivity of the RRS, which will make them faced with the failure of focus over edge subcarriers.
In the following, based on the delayed-RRS structure, we propose a new beamforming scheme that considers the RRS's frequency selectivity in the wideband near-field communication scenario.

\section{Min-rate Maximization Problem Formulation and Decomposition}

In this section, we propose a new optimization problem to improve the performance of our proposed system. However, due to the no-convex constraints and the coupled variabels of the problem, it is hard to solve it directly. Therefore, we decompose the problem into three subproblems.

\subsection{Max-min Rate Problem Formulation}

We aim to improve the sum rate of the considered system by designing the digital beamformer $\mathbf{g}_{m,i}$,the center phase shift $\theta^c_{i}$, and the time delay $\tau_i$ with ensuring the fairness among the multiple users. 
Therefore, we construct an optimization problem to maximize the minimum rate among multiple users by jointly optimizing $\mathbf{g}_{m,i},\theta^c_{i}$, and $\tau_i$~\cite{deng}. The optimization problem can be formulated as\vspace{-0.4cm}

\begin{subequations}
    \begin{align}
        (P1): &\operatorname*{maxmize}_{\mathbf{g}_{m,i},\theta^c_{i},\tau_i}\ \operatorname*{\min}_k	R_k , \label{obj1}\\
        \mathrm{s.t.} \quad 
        & \sum_{m=1}^{M}\sum_{i=1}^{K}\mathbf{g}^{H}_{m,i}\mathbf{g}_{m,i} \leq P, \label{con0} \\
        & \psi_{m,i}=e^{-j2\pi f_m\tau_i}\psi(\theta_i^c,f_m),\ \forall m,i,\label{con1}\\
        & \theta_i^c\in[0,2\pi],\ \forall i,\label{con2}\\
        & \tau_i\in[0,\tau_{\max}],\ \forall i.\label{con3}\\[-0.6cm]\notag
    \end{align}    
\end{subequations}
Constraint \eqref{con0} restricts the power consumption of the feeds. Constraint \eqref{con1} characterizes the frequency selectivity of RRS. Constraint \eqref{con2} and \eqref{con3} limit the available range of the center phase shift and time delay of each RRS element. 
Specifically, $P$ denotes the maximum power at the feeds. $\psi_{m,i}$ represents the response of the $i^{th}$ RRS element over the $m^{th}$ subcarrier. And $\tau_{\max}$ denotes the maximum delay of the time-delay units.

\subsection{Max-min Optimization Problem Decomposition}
\vspace{-0.1cm}
Note that problem (P1) is challenging because it is non-convex and the optimization variables are coupled. 
To reduce the complexity of the problem and decouple the three optimization variables, we decompose it into three subproblems, which are the digital beamforming, the analog beamforming, and the time-delay compensation respectively.

1) Digital Beamforming: Given the RRS configuration $\theta_i^c$ and the delay $\tau_i$, the digital beamforming subproblem can be written by 

\vspace{-0.7cm}

    \begin{align}
        \vspace{-0.2cm}
        (P2): &\operatorname*{maxmize}_{\mathbf{g}_{m,i}}\ \operatorname*{\min}_k	R_k , \hspace{0cm}\label{obj2}\\
        \mathrm{s.t.} \quad 
        & \eqref{con0}. \notag
        \vspace{-0.3cm}
    \end{align}

\vspace{-0.1cm}2) Analog Beamforming: Given $\mathbf{g}_{m,i}$ and $\tau_i$, we can design $\theta_i^c$ to perform the analog beamforming. 
The subproblem can be written as

\vspace{-0.7cm}

    \begin{align}
        \vspace{-0.3cm}
        (P3): &\operatorname*{maxmize}_{\theta_i^c}\ \operatorname*{\min}_k	R_k , \label{obj3}\\
        \mathrm{s.t.} \quad         
        & \eqref{con1},\eqref{con2}. \notag
        \vspace{-0.2cm}
    \end{align}

\vspace{-0.1cm}3) Time-Delay Compensation: Given $\mathbf{g}_{m,i}$ and $\theta_i^c$, we can optimize $\tau_i$ to compensate for the beam split problem mentioned above. The subproblem can be written as

\vspace{-0.45cm}

    \begin{align}
        \vspace{-0.2cm}
        (P4): &\operatorname*{maxmize}_{\tau_i}\ \operatorname*{\min}_k	R_k , \label{obj4}\\
        \mathrm{s.t.} \quad 
        & \eqref{con1},\eqref{con3}. \notag
        \vspace{-0.3cm}
    \end{align}   

\section{Joint Min-rate Optimization Algorithm Design}
Then we will develop a minimum rate maximization algorithm to obtain a suboptimal solution of (P1) by iteratively solving the subproblem (P2), (P3), and (P4).

\subsection{Digital Beamforming Algorithm}

Note that the objective function of the digital beamforming subproblem (P2) is the minimum of a series of rates, which is discontinuous. Through introducing a auxiliary variable $\eta$, we transform the discontinuous objective function into a continuous objective function which is easier to handle. We can reformulate (P2) as (P2a) accordingly. 

\begin{subequations}
    \begin{align}
        (P2a): 
        &\operatorname*{maxmize}_{\eta , \mathbf{g}_{m,i}} \eta, \\
        \mathrm{s.t.} \quad &\sum_{m=1}^{M}\log_2 \left( 1+\gamma_{m,k}\right) \geq \eta,\ \forall k,\ \label{p2acon1}  \\
        & \sum_{m=1}^M \sum_{i=1}^{K}\mathbf{g}_{m,i}^H\mathbf{g}_{m,i} \leq P, \label{p2acon2}
    \end{align}    
\end{subequations}
where $\eta$ represents the minimum sum rate among the $K$ users.

It is easy to verify that problem (P2a) is equivalent to problem (P2). The constraint on the auxiliary variable $\eta$ \eqref{p2acon1} is derived from the original objective function \eqref{obj2}. Noting that it is a non-convex constraint, we handle it using semidefinite relaxation (SDR) and successive convex approximation (SCA). Firstly, we apply SDR to convert the quadratic transform in the calculation of $\gamma_{m,k}$ \eqref{sinr} into affine transform. Considering $\mathbf{R}_{m,k}=\mathbf{B}_m^H\bm{\Psi}_m^H\mathbf{h}_{m,k}^*\mathbf{h}_{m,k}^T\bm{\Psi}_m\mathbf{B}_m$ and $\mathbf{V}_{m,i}=\mathbf{g}_{m,i}\mathbf{g}_{m,i}^H$, we have
\begin{align}
    \gamma_{m,k} = \frac{\mathrm{Tr}(\mathbf{R}_{m,k}\mathbf{V}_{m,k})}{\sigma^2+\sum_{i\neq k} \mathrm{Tr}(\mathbf{R}_{m,k}\mathbf{V}_{m,i})},\\
    \sum_{m=1}^M \sum_{i=1}^{K}\operatorname{trace}\left(\mathbf{V}_{m,i}\right) \leq P. \label{p2bcon}
\end{align}
Where the $\mathbf{V}_{m,i}$ is our new optimization variable, which is a semi-positive definite matrix and needs to satisfy
\begin{align}
    &\mathbf{V}_{m,i}\succeq 0, \label{sedi1}\\
    &rank(\mathbf{V}_{m,i})=1.\quad \label{rk1}
\end{align}
Here we relax the rank-one constraint \eqref{rk1} since it is non-convex and hard to handle. Then we use SCA to cope with the non-convex constraint \eqref{p2acon1} and solve for $\mathbf{V}_{m,i}$. Noting that \eqref{p2acon1} is equivalent to 
\begin{small}
\begin{equation}
    \begin{aligned}
        \forall k: &\sum_{m=1}^{M}\left[\log_2 \left( \sigma^2 +\sum_{i=1}^K \mathrm{Tr}(\mathbf{R}_{m,k}\mathbf{V}_{m,i})\right)\right.\\
        &\left.-\log_2\left( \sigma^2+\sum_{i=1,i\neq k}^K \mathrm{Tr}(\mathbf{R}_{m,k}\mathbf{V}_{m,i})\right)\right] \geq \eta. \label{ecop2a}
    \end{aligned}
\end{equation}
\end{small}It is still non-convex. We further relax it by applying the first-order Taylor expansion to its second log term. We can derive a inequality constraint from \eqref{ecop2a} as follows

\begin{equation}
    \begin{aligned}
        \\[-1.1cm]\\
        \hspace{-0.4cm}
        \forall k: &\sum_{m=1}^M\left[\log_2 \left( \sigma^2 +\sum_{i=1}^K \mathrm{Tr}(\mathbf{R}_{m,k}\mathbf{V}_{m,i})\right)- \right.\\
        &\quad\ \ \ a_{m,k}\left(\sum_{i=1,i\neq k}^K \mathrm{Tr}(\mathbf{R}_{m,k}\mathbf{V}_{m,i})-\right.\\
        &\quad\left.\left.\sum_{i=1,i\neq k}^K \mathrm{Tr}(\mathbf{R}_{m,k}\mathbf{V}^0_{m,i})\right)-b_{m,k}\right] -\eta \geq 0,\hspace{-0.3cm}\label{con2b}
    \end{aligned}
\end{equation}
where $\mathbf{V}^0_{m,i}$ denotes a solution of $\mathbf{V}_{m,i}$ in the last iteration. $a_{m,k}=\frac{1}{\ln2(\sigma^2+\sum_{i=1,i\neq k}^K \mathrm{Tr}(\mathbf{R}_{m,k}\mathbf{V}^0_{m,i}))}$ and $b_{m,k}=\log_2(\sigma^2+\sum_{i=1,i\neq k}^K \mathrm{Tr}(\mathbf{R}_{m,k}\mathbf{V}^0_{m,i}))$ denote the slope and the intercept of the first order Taylor expansion of $\log_2 ( \sigma^2 +\sum_{i=1,i\neq k}^K \mathrm{Tr}(\mathbf{R}_{m,k}\mathbf{V}_{m,i}))$ at $\mathbf{V}^0_{m,i},\forall i$. Note that this constraint is convex, and any solution $\mathbf{V}_{m,i}$ that satisfies this constraint will also satisfy constraint~\eqref{ecop2a}. In this way, we can derive a convex optimization problem (P2b) from (P2a).
\begin{align}
	(P2b):
	&\operatorname*{maxmize}_{\eta , \mathbf{V}_{m,i}}\ \eta, \hspace*{0.9cm}\\
	\mathrm{s.t.} \quad &\eqref{p2bcon}\eqref{sedi1}\eqref{con2b}.\notag
\end{align}

The problem can be optimally solved by existing convex optimization solvers, such as CVX~\cite{2014cvx}. Alternately, we find the optimal $\mathbf{V}_{m,i}$ for (P2b) and use the solution to update the $\mathbf{V}^0_{m,i}$. Repeating until convergence, we can achieve a suboptimal solution for $\mathbf{V}_{m,i}$ under the original constraint~\eqref{ecop2a}. 

Finally, since we relax the constraint $rank(\mathbf{V}_{m,i})=1$, the Gaussian randomization technique is used for obtain a feasible solution $\mathbf{g}_{m,i}$ for (P2a) from $\mathbf{V}_{m,i}$~\cite{wuqingqing1}. It's important to note that, during the whole SCA process, we consistently take $\mathbf{V}_{m,i}$ as the optimization variable instead of $\mathbf{g}_{m,i}$. This helps avoid the performance degradation when recovering $\mathbf{g}_{m,i}$, as the rank of
$\mathbf{V}_{m,i}$ may not necessarily be 1. until the final stage, where $\mathbf{V}_{m,i}$ is restored to $\mathbf{g}_{m,i}$. In this way, we can ensure the monotonic increment of the optimization objective $\eta$ throughout the entire digital beamforming subproblem, thereby guaranteeing the stability and convergence of the training procedure. 

The details of the proposed algorithm are summarized in \textbf{Algorithm~\ref{digital}}.

\begin{figure}[t]
    \centering
\vspace{-0.2cm}
\begin{minipage}{1\linewidth}
\begin{algorithm}[H]
    \renewcommand{\algorithmicrequire}{\textbf{Input:}}
	\renewcommand{\algorithmicensure}{\textbf{Output:}}
	\caption{Digital Beamforming}\label{digital}
    \begin{algorithmic}[1] 
        \REQUIRE $\mathbf{B}_m$,$\bm{\Psi}_{m}$ and $\mathbf{h}_{m,i}$; 
        \ENSURE $\mathbf{g}_{m,i}$; 
		\STATE initialize $\mathbf{g}_{m,i}$;
        \STATE calculate $\mathbf{V}_{m,i}^0$ from $\mathbf{g}_{m,i}$;
        \REPEAT
			\STATE calculate $a_{m,k}$ and $b_{m,k}$ from $\mathbf{V}_{m,i}^0$;
			\STATE $\mathbf{V}_{m,i}=solve(P2b)$;
			\STATE update $\mathbf{V}_{m,i}^0$;
		\UNTIL{convergence}
        \STATE obtain $\mathbf{g}_{m,i}$ from $\mathbf{V}_{m,i}^0$ with Gaussian randomization;
        \STATE \textbf{return} $\mathbf{g}_{m,i}$.
    \end{algorithmic}
\end{algorithm}
\end{minipage}
\vspace{-0.4cm}
\end{figure}

\subsection{Analog Beamforming Algorithm}

In this part, we introduce an algorithm based on the barrier function to solve the analog beamforming subproblems.

When $\mathbf{g}_{m,i}$ and $\tau_i$ is fixed, we optimize $\theta_i^c$ through solving the problem (P3). Similarly, we turn it into a maximum problem with continuous objective function by introducing a auxiliary variable $\eta$, which represents the minimum sum rate among $K$ users. Then we can reformulate P3 as P3a

\begin{subequations}
\begin{align}
	(P3a): 
	&\operatorname*{maxmize}_{\eta , \theta_i^c}\ \eta, \hspace*{1.5cm}\\
	\mathrm{s.t.} \quad & 
    \sum_{m=1}^{M}\log_2 \left( 1+\gamma_{m,k}\right) \geq \eta,\ \forall k,\label{p3acon1}  \\
    & \psi_{m,i}=e^{-j2\pi f_m\tau_i}\psi(\theta_i^c,f_m),\ \forall m,i,\label{p3ac2}\\
    & \theta_i^c\in[0,2\pi],\ \forall i,\label{p3ac3}
\end{align}    
\end{subequations}

We use barrier method to cope with the derived non-convex constraint~\eqref{p3acon1}. Specifically, we introduce a barrier function in the objective function to penalize the deviation from the constraint~\eqref{p3acon1} severely, and then we can relax the undesired complex constraint and approximate the suboptimal solution. 
The barrier function $u_k(\cdot)$ is defined as
\begin{equation}
u_k(\eta , \boldsymbol{\theta}) =-e^{\lambda_k\left(\sum_{m=1}^{M}\log_2 \left( 1+\gamma_{m,k}\right)-\eta\right)}, 
\end{equation}
where $\boldsymbol{\theta}$ denotes $\left[\theta_1,\cdots,\theta_{N_R}\right]$ and $\lambda_k$ denotes the weight coefficient of the barrier function. When we deviate from constraint (2a), the $u_k(\cdot)$ would drop sharply.
This guides the formulation of the new optimization problem (P3b) from (p3a) as follows
\begin{align}
	(P3b): 
	&\operatorname*{maxmize}_{\eta , \theta_i^c}\  \eta + \sum_{k=1}^{K}u_k(\eta , \boldsymbol{\theta}),\quad\\
	\mathrm{s.t.} \quad &\eqref{p3ac2}\eqref{p3ac3}. \notag
    \vspace{-0.2cm}
\end{align}
The problem has a very simple constraint structure, which brings convenience to the optimization process. We use the gradient descent method to solve for a suboptimal solution of (P3b). After that, we increase $\lambda_k$ gradually, decrease the search step size and solve the (P3b) iteratively. Note that as the value of $\lambda_k$ increase, the deviation will bring greater degradation to $u_k(\cdot)$. So when $\lambda_k$ is sufficiently large, the solution of (P3b) will match the constraint \eqref{p3acon1} within a small error margin $\epsilon$, i.e., $\sum_{m=1}^{M}\log_2 \left( 1+\gamma_{m,k}\right) \geq \eta -\epsilon,\forall k$ will be satisfied. Finally, we can find a local optimal solution of (P3a) within the certain error tolerance.

Note that since we use e-exponential barrier function, we do not need to guarantee that the starting point of each iteration is within the feasible region. It will naturally converge to an approximately feasible solution. In addition, by choosing the step size carefully, we can avoid the solution from crossing the local optimal point and falling into another local optimal point. This can guarantee the convergence of the iterations.

There are two reasons why we no longer use the SDR method to solve the problem. Firstly, the number of RRS elements is much larger than the number of feed antennas, which means the dimension of the solution vector of the problem is much greater than the former. On one hand, it would take much longer time to reach the solution. On the other hand, when we compress the rank of a high dimensional semidefinite matrix to 1, we introduce a larger error. Secondly, the constraint \eqref{p3ac2} is too complex and irregular, which would lead to unbearable losses of performance when we try to restore a feasible $\boldsymbol{\Psi}_m$ from a semi-positive definite matrix. Therefore, we need to find a new method to solve the problem.

\subsection{Time-Delay Compensation Algorithm}

Since the problems (P3) and (P4) have very similar forms, we can use similar methods to solve (P4). The only difference is that, in the (P4), the variable $\tau_i$ needs extra constraint because we cannot choose arbitrarily large time delays. 

Similarly, we introduce a auxiliary variable $\eta$ to represent the minimum sum rate among $K$ users and construct a barrier function $v_k(\eta,\boldsymbol{\tau})$ as
\begin{equation}
    v_k(\eta , \boldsymbol{\tau}) =-e^{\lambda_k\left(\sum_{m=1}^{M}\log_2 \left( 1+\gamma_{m,k}\right)-\eta\right)},
\end{equation}
where $\boldsymbol{\tau}$ denotes $\left[\tau_1,\cdots,\tau_{N_R}\right]$ and $\lambda_k$ denotes the weight coefficient of the barrier function. We can reformulate (P4) as (P4a) as follows
\begin{subequations}
    \begin{align}
        (P4a): 
        &\operatorname*{maxmize}_{\eta , \tau_{i}}\ \eta +\sum_{k=1}^Kv_k(\eta,\boldsymbol{\tau}), \hspace*{1.5cm}\\
        \mathrm{s.t.} \quad & 
        \psi_{m,i}=e^{-j2\pi f_m\tau_i}\psi(\theta_i^c,f_m),\ \forall m,i,\hspace{-0.4cm}\label{p4ac2}\\
        & \tau\in[0,\tau_{max}],\ \forall i,\label{p4ac3}
    \end{align}    
\end{subequations}
Here we set $\tau_{max} = \frac{1}{f_c}$ to enable the time-delay units to provide an maximum phase shift approximately equal to $2\pi$ for the signals. Then, we can iteratively use the gradient descent method to solve for the local optimal solution of $\tau_i$. Here we omit the details, as the process is similar to the solving for  $\theta_i^c$.

\subsection{Overall Algorithm Description}

Based on the discussion above, we propose an iterative algorithm to solve the (P1) by optimizing the $\mathbf{g}_{m,i}$, $\theta_i^c$ and $\tau_i$ jointly. We first initialize the three variables randomly, and then iteratitively solve the subproblems (P2), (P3), and (P4) to update the variables.  The algorithm can be summarized as \textbf{Algorithm~\ref{algorithmall}}.

\begin{figure}[t]
    \centering
\vspace{-0.2cm}
\begin{minipage}{1\linewidth}
\begin{algorithm}[H]
    \renewcommand{\algorithmicrequire}{\textbf{Input:}}
	\renewcommand{\algorithmicensure}{\textbf{Output:}}
	\caption{Min-rate Optimization Algorithm}
    \label{algorithmall}
    \begin{algorithmic}[1] 
        \REQUIRE  $\mathbf{B}_m$ and $\mathbf{h}_{m,i}$; 
	    \ENSURE $\mathbf{g}_{m,i}$, $\theta_i^c$ and $\tau_i$; 
        \STATE Randomly Initialize $\mathbf{g}_{m,i}$, $\theta_i^c$ and $\tau_i$;
		\FOR{iter = 1 to \textit{iter}}
            \STATE calculate $\mathbf{V}_{m,i}^0$ from $\mathbf{g}_{m,i}$;
            \REPEAT
                \STATE solve (P2b) to update $\mathbf{V}_{m,i}^0$;
            \UNTIL{convergence}
			
			\STATE update $\mathbf{g}_{m,i}$ from $\mathbf{V}_{m,i}$;
			\STATE solve (P3) to update $\theta_i^c$;
            \STATE solve (P4) to update $\tau_i$;
		\ENDFOR{}
        \STATE \textbf{return} $\mathbf{g}_{m,i}$, $\theta_i^c$ and $\tau_i$;
    \end{algorithmic}
\end{algorithm}
\end{minipage}
\vspace{-0.4cm}
\end{figure}

\section{Performance Analysis}
\subsection{Algorithm Convergence}
First, in the digital beamforming, solving the convex optimization problem (P2b) ensures that the optimization objective $\eta$ increases monotonically during iterations. Besides, $\eta$ represents the user's communication rate, which is evidently upper-bounded. Therefore, the iterative process in {Algorithm~\ref{digital}} is guaranteed to converge. 

Note that when recovering the beamformers $\mathbf{g}_{m,i}$ from the semi-positive definite matrixs $\mathbf{V}_{m,i}$, $\eta$ may slightly decrease.  To ensure the strict increase of the minimum rate in Algorithm~\ref{digital}, we introduce a checkpoint after the recovering to verify whether the newly obtained digital beamformers can improve the minimum rate. If not, we will discard the new solution and restore the previous one. 

Second, in the analog beamforming, for any fixed $\lambda_k$, the optimization objective $\eta$ is guaranteed to increase monotonically during the gradient descent. 
As $\eta$ is upper-bounded, the optimization of the problem (P3b) is guaranteed to converge, and so is the problem (P3a) because (P3a) is solved by solving a finite number of instances of problem (P3b). Problem (P4a) also leads to similar conclusions. 

Note that we cannot guarantee the strict increase of the optimization objective $\eta$ in (P3a). This is because the barrier function introduced in (P3b) may result in small errors or cause the optimization objective to jump from one local optimum to another. However, by setting an appropriate initial value for  $\lambda_k$, we can at least ensure that the optimization objective of (P3a), i.e., the minimum user rate, increases strictly within the allowable error range.

Therefore, since the minimum user rate can be ensured to increases monotonically in each subproblem and it is evidently upper-bounded, we can guarantee the convergence of the entire algorithm within the allowable error range.

\subsection{Algorithm Complexity}

To estimate the time cost of our proposed algorithm, we analyze the complexity of the three sub-problems separately and then present the complexity of each iteration of the overall optimization algorithm.

\begin{itemize}
    \item \textbf{Digital beamforming algorithm}: 
    In the solving of the problem (P2a), let $L$ denote the number of iterations for SCA. For each SCA iteration, the convex optimization solver uses the interior-point method (IPM) to solve the convex optimization problem (P2b). We use the complexity $\mathcal{O}(\sqrt{N}\ln(1/\epsilon))$ in the Short-Step Path-Following Algorithm to estimate the number of iterations required by the IPM, and use $\mathcal{O}(MKN^3)$ to estimate the time complexity of each iteration, where $MK$ implys the number of the constraints and $N^3$ denotes the time complexity of the matrix multiplication and matrix inversion. The overall complexity of the digital beamforming algorithm is $\mathcal{O}(LMKN^{3.5}\ln(1/\epsilon))$.

    \item \textbf{Analog beamforming algorithm}: 
    We use the gradient descent method to solve the problem (P3b). The complexity of calculating the gradients of the objective function with respect to $\eta$ and $\theta_i^c$ is approximately $\mathcal{O}(MKN_R^3)$, where $MK$ implys the number of the constraints and $N_R^3$ corresponds to the complexity of computing the vector gradient. Besides, the step size decreases exponentially during the descent process, and the weight coefficients of the barrier function increase exponentially too. So we use $\mathcal{O}(\ln(1/\epsilon))$ to estimate the iterations of the descent method in (P3b) and the barrier method in (P3a), where $\epsilon$ represents the desired accuracy of the solution. The overall complexity of the analog beamforming algorithm is $\mathcal{O}(MKN_R^3\ln(1/\epsilon)^2)$.
\end{itemize}

Since the problem (P4a) is solved similarly to the problem (P3a), the complexity of the time-delay compensation algorithm is $\mathcal{O}(MKN_R^3\ln(1/\epsilon)^2)$ too. The overall complexity of each iteration of the proposed algorithm is $\mathcal{O}(LMKN^{3.5}\ln(1/\epsilon)+MKN_R^3\ln(1/\epsilon)^2)$.

\subsection{Optimal Distance between the Feeds and the RRS}

To evaluate the impact of the distance between the feeds and the RRS on the performance of the base station, we consider a single-feed base station 
and assume that through the analog beamforming at the RRS, the electromagnetic waves radiated onto the RRS can be perfectly focused on the user through the RRS. Then the ideal amplitude gain can be written as
\begin{equation}
    A = \sum\sqrt{\frac{G_t F_rS_0}{4\pi d^2}}.
\end{equation}
The expression sums over all RRS elements, with$G_t$, $F_r$ and $d$ varying across different RRS elements, where $S_0$ denotes the receiving area of the RRS elements, $G_t$ denotes the transmit gain pattern of the feeds, $F_r$ represents the normalized radiation pattern of the RRS elements, the $d$ denotes the distance between the feed antenna and the RRS element and $\frac{1}{d^2}$ represents the attenuation of electromagnetic waves as they propagate through the air. As the distance increases, the attenuation will increase, but a larger part of the RRS will fall within the region of concentrated radiation from the feed, meaning the directivity factor $F_r$ and $G_t$ will increase too. Therefore we have the Proposition 1 below, which is proved in Appendix A.

\textit{Proposition 1:} In the case of a single-feed RRS-enabled base station, assuming that the RRS can focus the beams on the user perfectly, there exists an optimal distance between the feed and the RRS when the directivity of the feed and the RRS is large enough.

However, under the wideband communication scenario, it is impossible for the RRS to perfectly focus the beams over each subcarrier. Based on this point, we assume that the arriving signal of each path has a random phase shift $\varphi$, which denotes the uncontrollable errors brought by the frequency selectivity. The amplitude gain can be rewriten as
\begin{equation}
    A^\prime = \sum\sqrt{\frac{G_t F_r S_0}{4\pi d^2}}e^{j\varphi}.
\end{equation}  
If we treat all variables as random variables, $\phi$ is independent of the other variables. Therefore, we have
\begin{equation}
    A^\prime = \mathbb{E}(e^{j\varphi})A,
\end{equation}
where $\mathbb{E}(e^{j\varphi})$ represents the impact of frequency selectivity on the gain.

This proposition is also insightful for the multi-feed and multi-user scenario, which is confirmed in the simulation results presented in the next section.

\section{Simulation Results}

In this section, we evaluate the performance of our proposed algorithm in terms of the achieved minimum rate among multiple users. We show how the system influenced by the SNR, the frequency selectivity of the RRS and the number of quantization bits for discrete phase shifts. For comparison, the following schemes are introduced as well.
\begin{itemize}
    \item \textbf{Without TD}: This method follows the same procedure as our proposed method, except the optimization of $\tau_i$ (P4) and sets all $\tau_i$ to zero.
    \item \textbf{Ignoring Frequency Selectivity of RRS}: This method neglects the frequency selectivity of the RRS when solving out the beamformers.
    \item \textbf{Random Configuration}: This method sets each $\tau_i$ to zero and makes each $\theta_i^c$ randomly and uniformly distributed within $\left[0,2\pi\right]$, then solves (P2) to obtain the digital beamformer only. 
    \item \textbf{Far-field Wave Model}: This method uses far-field plane wave model to estimate the channel when performing the beamforming with our proposed method, but calculate the achieved minimum rate with using accurate spherical wave model. 
\end{itemize}

\subsection{Simulation Setup}

\begin{table}
    \begin{center}
    \caption{Simulation parameters}
    \label{tab1}
    \begin{tabular}{| c | c |}
    \hline
    \textbf{Parameters} & \textbf{Values}\\
    \hline
    \hline
    Central carrier frequency $f_c$  & 30 GHz\\
    
    \hline
    Bandwidth of the system $W$ & 1 GHz\\
    \hline
    Number of subcarriers $M$ & 16\\
    \hline
    Number of RRS elements $N_R$ & 81\\
    \hline 
    Number of feeds $N$ & 4\\
    \hline 
    Number of users $K$ & 2\\
    \hline
    Transmit power $P$ & 10 dB\\
    \hline 
    Variance of the AWGN $\sigma^2$ & -3 dB\\
    \hline
    The distance between the feeds and the RRS & 0.1 m\\
    \hline
    The frequency selectivity of RRS $k_p$ & 1\\
    \hline 
    Number of iterations \textit{iter}  & 3\\
    \hline
    \hline 
    \end{tabular}
    \end{center}
\end{table}
We consider a three-dimensional (3D) coordinate system. A uniform rectangular RRS is located in x-y plane, its center coincides with the origin, and its spacing is set to half center wavelength. The linear feeds are uniformly distributed in front of the RRS with a vertical distance of 0.1 m, which is parallel to the x-axis and heading towards the RRS. We set the distance between $K$ users and the RRS to about 0.25 m, which is within the Rayleigh distance of the RRS. The $\alpha_t$ in $G_t$ is set to $62$, which is often used in the X-band horn antenna, and the $\alpha_r$ in $F_r$ and $G_r$ are set to $3$. The receiving area of each RRS element is set to $S_0=0.25\ \mathrm{cm}^2$ and  $A_r$ is set to 1. The maximum delay of the RRS element is set to $\tau_{\max}=\frac{1}{2f_c}$. 

\begin{figure}[t]
    \vspace{0.35cm}
    \centerline{\includegraphics[width=8.7cm]{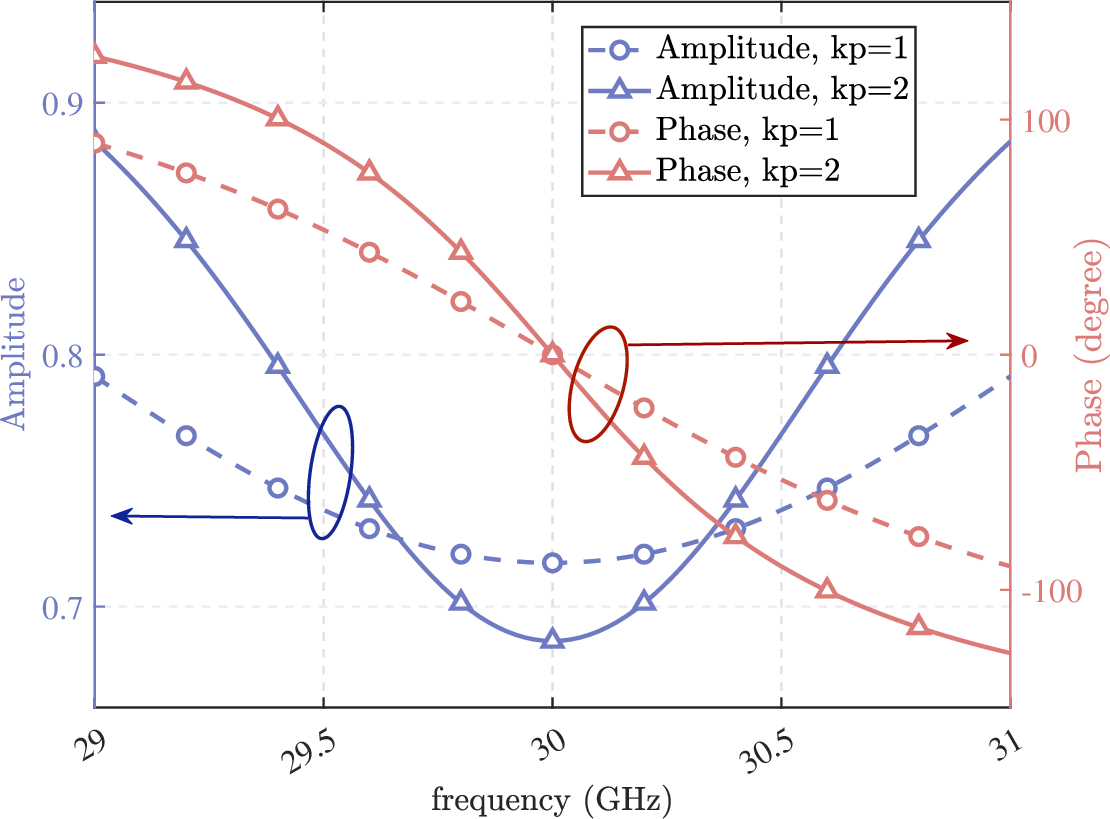}}
    \vspace{-0cm}
    \caption{The refractive coefficient of refractive surfaces varies with frequency.}
    \vspace{0cm}
    \label{rrsresponse}
\end{figure}

Fig.~\ref{rrsresponse} shows the modeling results of two sets of responses. Referring to the modeling in~\cite{practical2}, we construct two functions to express the amplitude-frequency and phase-frequency response of RRS elements respectively. Moreover, we introduce $k_p$ to represent the response's sensitivity to the frequency. In our modeling, the smaller the $k_p$ is, the closer the frequency response of the RRS is to that of ideal phase shifters, i.e., the phase shift and amplitude are the same at different frequencies when $k_p$ is 0. Other simulation parameters can be found in Table~\ref{tab1}

\subsection{Numerical Results}

\begin{figure}[t]
    \vspace{0.35cm}
    \centerline{\includegraphics[width=8cm]{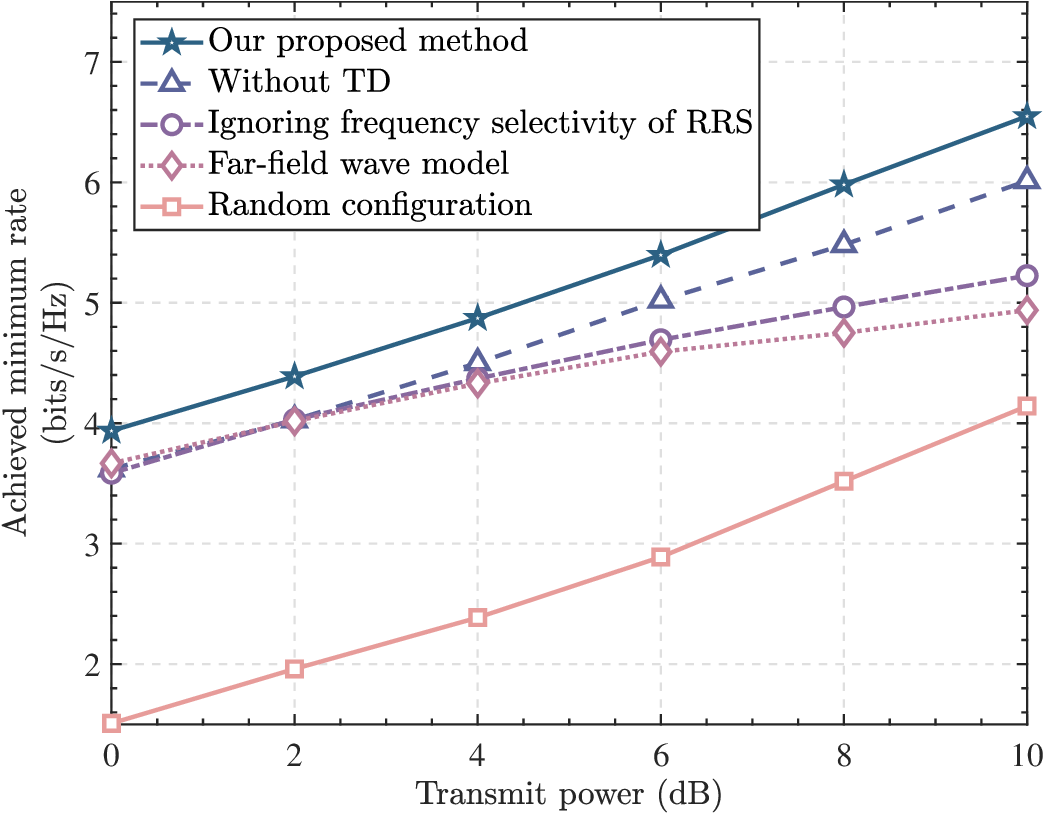}}
    \vspace{-0cm}
    \caption{Achieved minimum rate versus transmit power.}
    \vspace{0cm}
    \label{result1}
\end{figure}

Fig.~\ref{result1} shows the achieved minimum rate versus the transmit power. The variance of the noise is fixed and the transmit power is distributed between 0 dB and 10 dB. Since the base station provides more power resources, the minimum sum-rate increases with the transmit power. We can observe that our proposed method outperforms the other methods. Specifically, comparing the \textit{Our proposed method} and \textit{Without TD}, we can find that the compensation effect of TD can provide a gain equivalent to the improvement brought by an power increase of about 2 dB. Besides, ignoring the frequency selectivity of the RRS does lead to a noticeable performance loss, which is comparable to the loss of ignoring the near-field channel condition. These results indicate the effectiveness of our proposed method and stress the importance of considering the frequency selectivity of RRS and the near-field channel condition jointly. What's more, we can observe that the loss of ignoring the frequency selectivity of the RRS or the near-field channel condition becomes more significant as the transmit power is growing larger. This is because the two kind of loss mainly comes from the interference among different users' signals, which occurs during the transmission and will increase as the transmit power increases.

\begin{figure}[t]
    \vspace{0cm}
    \hspace{-0.2cm}
    \centerline{\includegraphics[width=8.9cm]{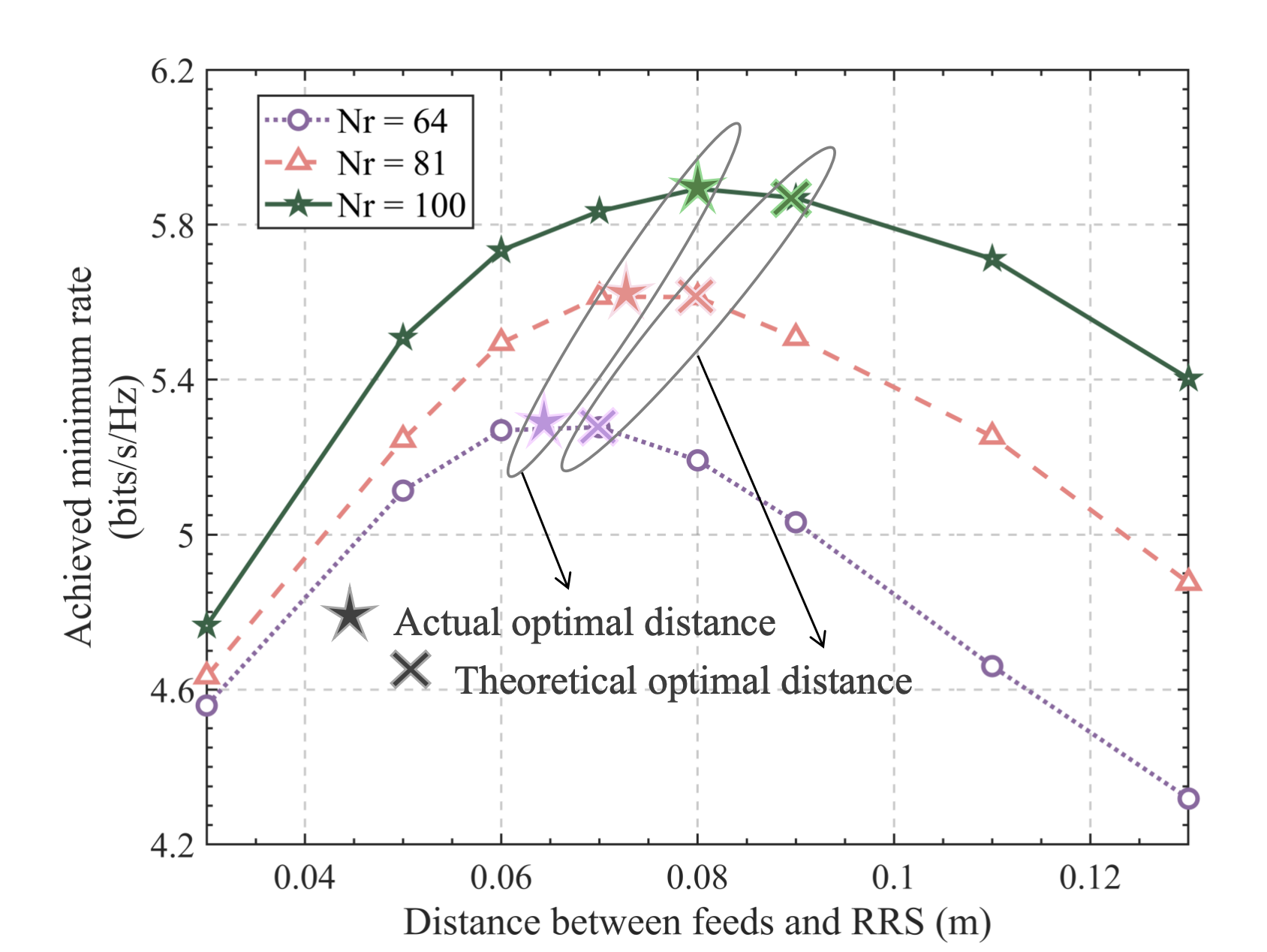}}
    \vspace{-0cm}
    \caption{Achieved minimum rate versus distance between feeds and RRS.}\label{result2}
    \vspace{-0.2cm}
\end{figure}

Fig.~\ref{result2} shows the achieved minimum rate versus the distance between the feeds and the RRS. Let the distance be distributed between 0.03 m and 0.12 m. And the number of RRS elements is set to 64, 81 and 100 respectively. 
The rest of the parameters are consistent with Fig.~\ref{result1}.
We can find that as the distance increases, the achieved minimum rate initially increases and then decreases. In addition, with the increasing number of elements in RRS, the optimal value of distance also increases. This is because an increase in distance leads to higher path loss, but it also increases the effective radiating area of the feed antenna on the RRS. The array gain would increase as the effective radiating area increases. 
When the distance is very small, the increase in antenna gain overwhelms the path loss as the distance increases. But when the distance is large, the effective radiating area is limited by the size of the RRS, and path loss becomes dominant. So there exists an optimal feed-RRS distance for the best data rate, and it increases with the size of the RRS.
The optimal distance calculated numerically from Appendix A is marked with a cross in the figure, and the actual optimal distance from the simulation experiments is also marked with a star in the figure. It can be observed that the two are very close and exhibit the same trend with respect to the variation in the number of the RRS elements. Since four feed antennas were used in the simulation, it is reasonable that the actual optimal distance is slightly smaller than the theoretical optimal distance.
This result confirms the proposition 1 proposed in the previous section.

\begin{figure}[t]
    \vspace{0.35cm}
    \centerline{\includegraphics[width=8cm]{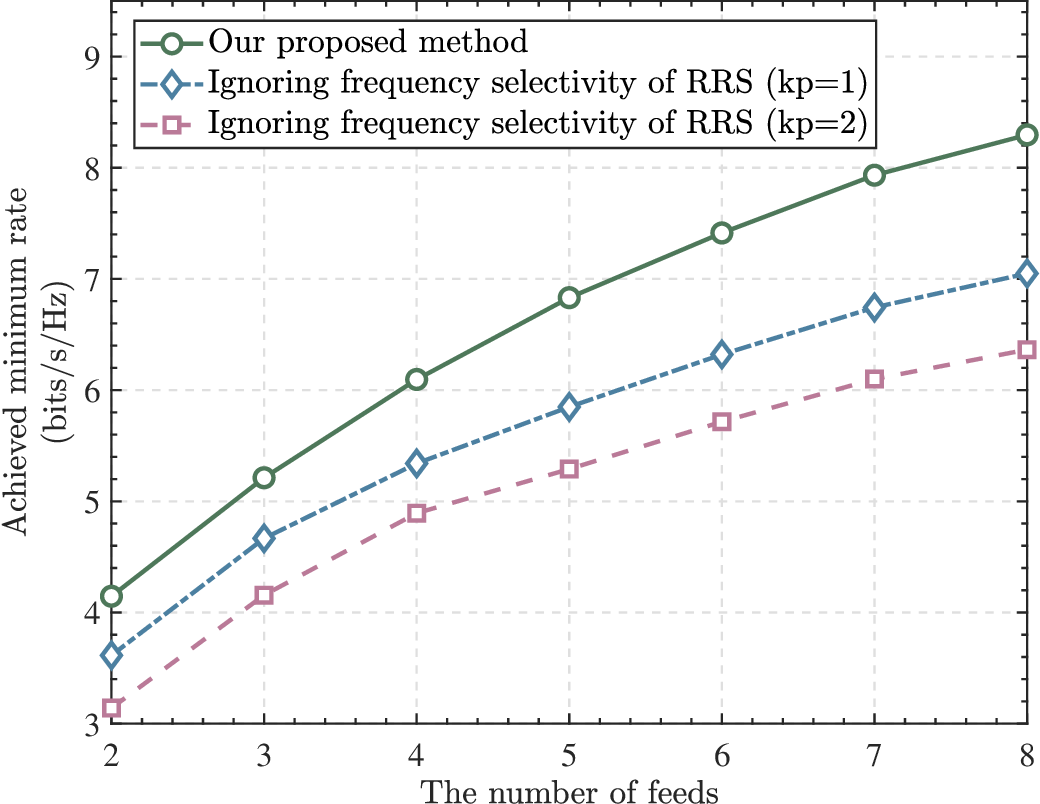}}
    \vspace{-0cm}
    \caption{Achieved minimum rate versus the number of the feeds.}
    \vspace{-0cm}
    \label{result3}
\end{figure}

Fig.~\ref{result3} shows the relationship between the achieved minimum data rate and the number of the feeds. We fix the transmit power to 10 db and set the number of RRS elements to $N_R=100$. The number of the users is set to $K=2$ and the number of the feeds $N$ ranges from 2 to 8. The $k_p$, which indicates the strength of the frequency selectivity of RRS, is set to $k_p=1$ in lines 1, 2, and $k_p=2$ in line 3. The rest of the parameters are consistent with Fig.~\ref{result1}. 
Generally, we can find that as the number of the feeds increases, the minimum rate increases too.  However, as the number of the feeds increases, the increase of the data rate will slow down. Besides, we can find that ignoring the frequency selectivity of the RRS element has a significant impact on the performance. As the frequency selectivity is stronger, the loss is more significant. This is because the stronger frequency selectivity of the RRS would lead to a larger error brought by ignoring the frequency selectivity, which would make the RRS harder to focus the beams and reduce the system's performance. At the same time, comparing the blue and orange lines, we can observe a phenomenon similar to Fig.~\ref{result1} that, when the number of feed antennas increases, the loss caused by ignoring the frequency selectivity of RRS also slightly increases. This indicates that more feed antennas can improve the power received by the RRS, but at the same time, it will lead to more interference from ignoring the frequency selectivity of the RRS.
\begin{figure}[t]
    \vspace{0.35cm}
    \centerline{\hspace{-0.13cm}\includegraphics[width=8.3cm]{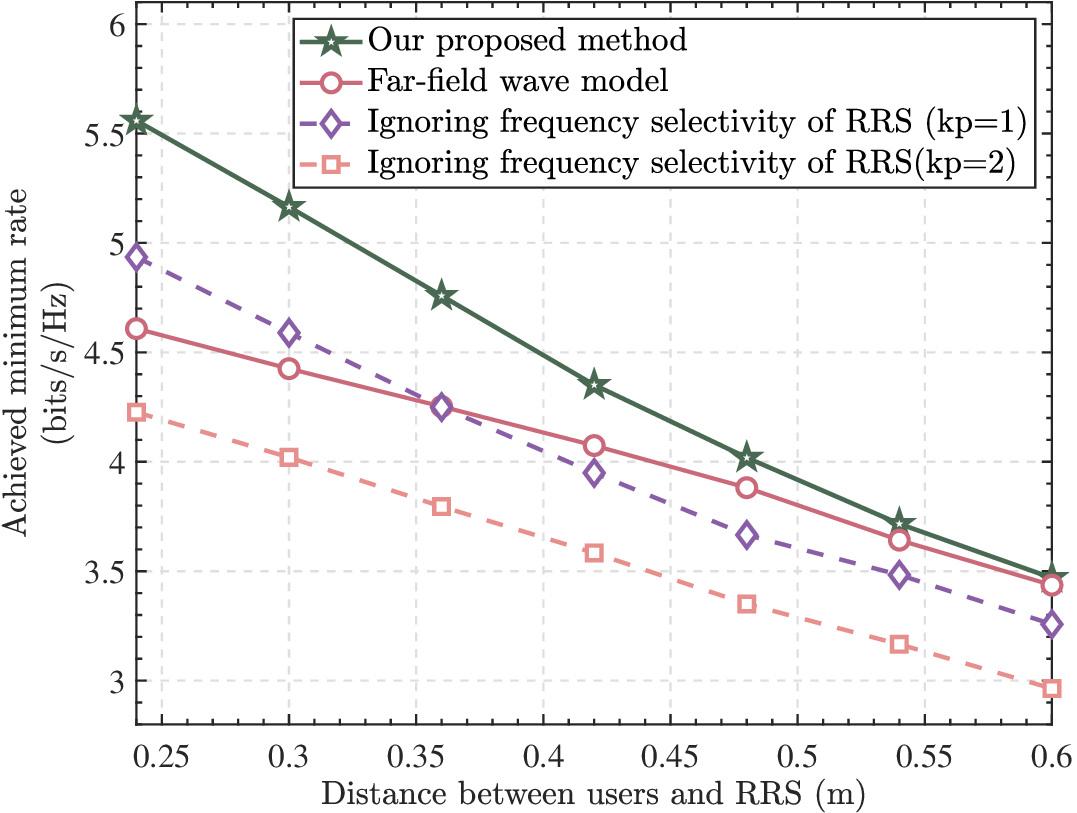}}
    \vspace{-0cm}
    \caption{Achieved minimum rate versus distance between RRS and users.}
    \vspace{0cm}
    \label{result5}
\end{figure}

Fig.~\ref{result5} shows the relationship between the achieved minimum rate and the distance between the users and the RRS.  We make the distance distributed between 0.24m and 0.6m. The $k_p$, which indicates the degree of frequency selectivity of the RRS, is set to $k_p=1$ in lines 1, 2, 3, and $k_p=2$ in line 4. It is obvious that as the distance increases, the achieved minimum rate decreases. This is because the path loss increases as the distance increases. Further, we focus on the losses in the three special cases compared to \textit{Our proposed method}. 
Comparing \textit{Our proposed method} and \textit{Far-field wave model}, we can find that, as the distance decreases, the loss brought by the plane wave model is increasing. This is because the plane wave model is only valid in the far-field region, where the near-field effect can be ignored. Specifically, when the distance is smaller than the Rayleigh distance, which is about 0.45m in this case, the loss is particularly significant. And from the other two instance, we can find that ignoring the frequency selectivity of the RRS element has a significant impact on the data rate performance. When the frequency selectivity gets stronger, the loss is more severe. Besides, there exists a critical distance: when the distance between the user and the BS exceeds this critical distance, the data rate loss caused by ignoring frequency selectivity is greater than that caused by ignoring the near-field effect, and vice versa. When the frequency selectivity is sufficiently strong, the critical distance becomes much smaller than the Rayleigh distance. In conclusion, simulation results show the necessity of considering the near-field effect and the frequency selectivity of RRS elements jointly, and when the users are not very close to the BS, it is even more important to consider the frequency selectivity of the RRS.

\begin{figure}[t]
    \vspace{-0cm}
    \centerline{\includegraphics[width=8cm]{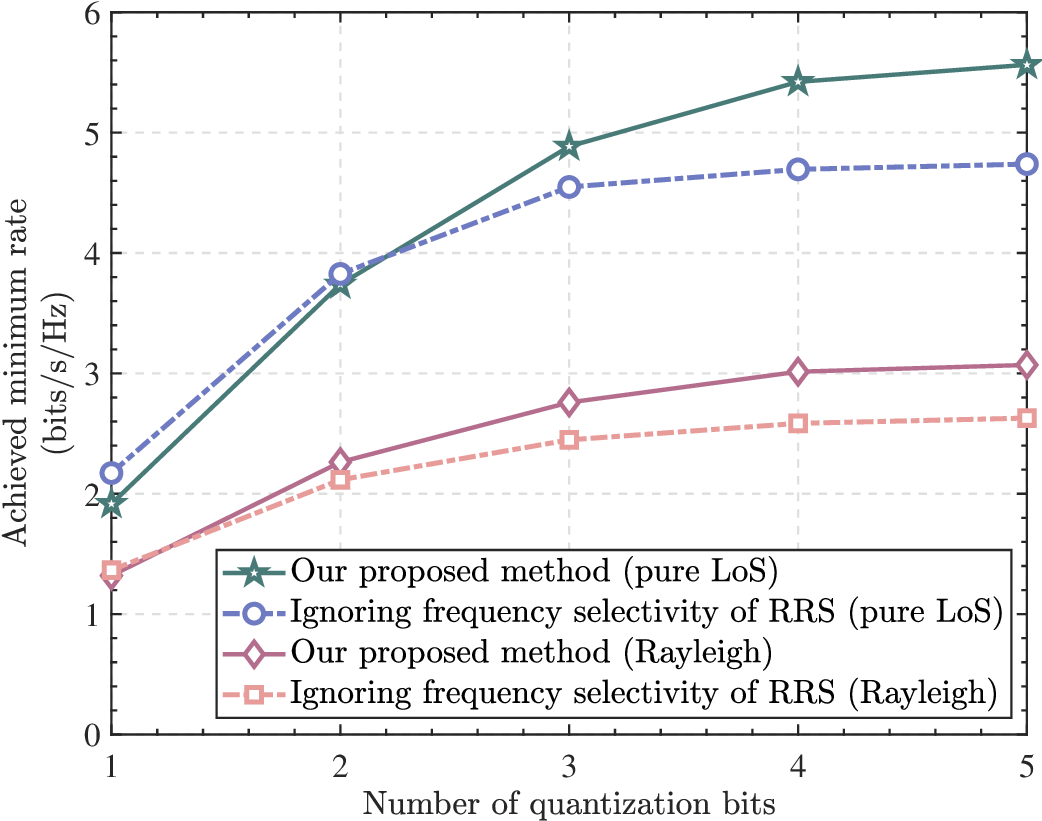}}
    \vspace{-0cm}
    \caption{Achieved minimum rate versus number of quantization bits.}
    \vspace{-0.4cm}
    \label{result4}
\end{figure}

Fig.~\ref{result4} depicts the achieved minimum user's rate versus the number of quantization bits. In each simulations, we keep only a finite number of phase shifts and convert the continuous configurations to discrete phase configurations based on the proximity principle. In the \textit{pure LoS} case, we only consider the LoS propagations from the RRS to the users during channel estimation. In the \textit{Rayleigh} case, we consider a more complex wireless channel environment, i.e., we use the Rayleigh channel model to estimate a rich-scattering NLoS channel. Let $\tilde{\mathbf{h}}_{m,i}$ denotes the Rayleigh channel vector, which can be given by
\begin{equation}
    [\tilde{\mathbf{h}}_{m,i}]_k = \left|[\mathbf{h}_{m,i}]_k\right|\hat{h}_{m,i,k},
\end{equation}
where $\hat{h}_{m,i,k} \sim \mathcal{C} \mathcal{N} (0,1)$ denotes the small-scale NLoS component. Note that we multiply the NLoS component by $\left|[\mathbf{h}_{m,i}]_k\right|$ to ensure that the expected power of the Rayleigh channel remains consistent with that of the LoS channel.
From the Fig.~\ref{result4}, we can observe that the achieved minimum rate increases with the number of quantization bits, both in \textit{pure los} and \textit{Rician} conditions. Moreover, when the number of quantization bits is small, the rate increases rapidly as the number of quantization bits increases, while finally slow down and converge to an upper bound when the number of quantilization bits is large. This result indicates that the quantization bits of the RRS's configuration have a significant impact on the system's performance and we can achieve a sufficiently large data rate with a not large number of quantization bits.
Meanwhile, as the number of quantization bits decrease, ignoring the frequency selectivity of the RRS cause less loss of achieved minimum rate, and even performs better. This is because ignoring the frequency selectivity during the optimization improves the system's tolerance to the errors introduced by quantizing the configurations of RRS. In addition, comparing the \textit{pure LoS} case and the \textit{Rician} case, we can find that the quantization of RRS configurations results in some data-rate loss in both channel environments. However, in the more complex channel environment, the RRS-based communication system performs worse.

\section{Conclusion}
In this paper, we have studied an RRS-enabled downlink wideband multi-user system, where a frequency selective RRS is adopted as the transmit antennas. The frequency selectivity would make the RRS harder to focus the beams and make the beam split effect more severe. Jointly considering the frequency selectivity of RRS and the near-field communication, we have designed an HBF scheme to maximize the minimum user's rate by optimizing the digital beamformer, the configuration of RRS and the delays of time-delay units alternately. Simulation results prove the effectiveness of our proposed scheme.

Specifically, several remarks can be drawn from the theoretical analysis and simulation results, providing insights into the RRS-enabled communication system. 
\begin{itemize}
    \item Compared to traditional phased arrays, the frequency selectivity of RRS causes the near-field beams to lose focus on the edge subcarriers, which would intensify the beam split problem and result in the performance degradation. 
    \item As the distance between the feeds and the RRS increases, the system's performance first increases and then decreases. There exists an optimal distance between the feeds and RRS, and the optimal distance would increase with the size of the RRS.
    \item Considering the finite quantization bits of the RRS's configuration, the minimum rate among the users would increase rapidly when the number of the quantization bits is small, and gradually converge to the ideal case where the configuration is continuously adjustable.

\end{itemize}

\vspace{-0.3cm}
\begin{appendices}
\section{The Optimal Feed-RRS Distance} 
We divide each RRS element into many small parts, each with an area of $dS$, and assume that the channel attenuation at each part is the same. Therefore, the ideal amplitude gain ${A}$ can be approximated in terms of integrals
\begin{equation}
    \begin{aligned}
        A &= \sum\sqrt{\frac{G_t F_rS_0}{4\pi d^2}} \\
        &= \int_{S}\sqrt{\frac{G_tF_r}{4\pi d^{2}S_0}}{dS}.
    \end{aligned}
\end{equation}
Where $S$ denotes the whole area of the RRS. $G_t = G(\alpha_t)$ denotes the transmit gain pattern of the feed according to~\eqref{radi_pattern}, $F_r$ that given by~\eqref{fr} denotes the normalized radiation pattern of RRS elements. The $d$ denotes the distance between the feed antennas and the integral area of RRS and $\frac{1}{d^2}$ represents the attenuation of electromagnetic waves as they propagate through the air. 

\begin{figure}[t]
    \vspace{-0.2cm}
    \centerline{\includegraphics[width=7cm]{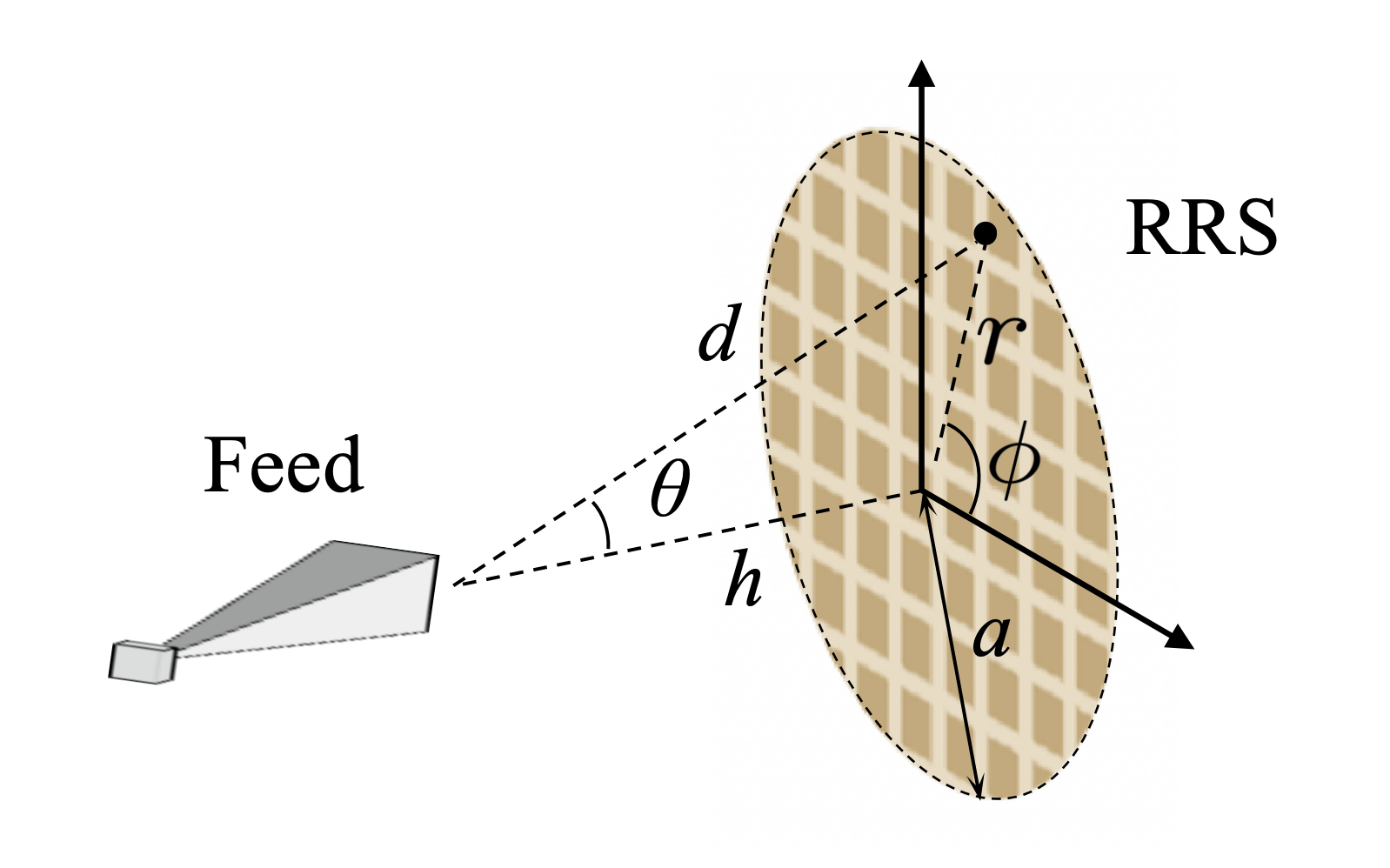}}
    \vspace{-0cm}
    \caption{A circle RRS illuminated by a feed antenna.}
    \vspace{-0.5cm}
    \label{appendix1}
\end{figure}

If we regard the RRS as a circle approximately, and establish a plane polar coordinate system on it, whose origin coincides with the center of the RRS. Then the surface integral can then be rewritten as a double integral.
\begin{equation}
    {A} = \int_\phi \int_r \sqrt{\frac{\alpha_t+1}{2\pi d^{2}S_0}}\cos^{\frac{\alpha_t+\alpha_r}{2}}\theta r drd\phi. \label{inte1}
\end{equation}
Here we substitute equation \eqref{radi_pattern} and \eqref{fr} into $G_t$ and $F_r$. The $\alpha_t$ and $\alpha_r$ describe the directivity of the feed antenna and the RRS elements. The meanings of $\theta$ and $\phi$ are illustrated in Fig.~\ref{appendix1}.

Furthermore, with the relation $ r = h \tan \theta$ and $d = \frac{h}{\cos \theta}$, we change the integral variable from $r$ to $\theta$. \eqref{inte1} can be written as a double integral
\begin{equation}
    {A} = \sqrt{\frac{\alpha_t+1}{2\pi S_0}}\int_\phi d\phi \int_\theta h \cos^{\frac{\alpha_t+\alpha_{tr}-4}{2}}\theta \sin \theta d\theta.
\end{equation}

The region of $\phi$ is $\left[0,2\pi\right]$, so $\int_{\phi}d\phi = 2\pi$. Letting the unconcerned term $\sqrt{\frac{\alpha_t+1}{2\pi S_0}}\int_\phi d\phi = C $ and $\cos\theta= x$, we have
\begin{equation}
    {A} = C\int_x h x^{\frac{\alpha_t+\alpha_{tr}-4}{2}}dx.
\end{equation}The region of $x$ is $\left[\sqrt{\frac{h^2}{a^2+h^2}},1\right]$. $a$ denotes the radius of the RRS, which is corresponding to the maximum of $\theta$ and the minimum of $\cos \theta$. The integral can be calculated as\newpage\noindent
\begin{equation}
    \begin{aligned}
        {A}&=C\frac{2h}{\alpha_{t}+\alpha_{tr}-2}\left[1-(\frac{h^{2}}{a^{2}+h^{2}})^{\frac{\alpha_{t}+\alpha_{tr}-2}{4}}\right]\\
        & = C\frac{2a}{(\alpha_{t}+\alpha_{tr}-2)\tan\theta_0}\left[1-\left(\sin\theta_0\right)^{\frac{\alpha_{t}+\alpha_{tr}-2}{2}}\right].
    \end{aligned}
\end{equation}
Here $\theta_0$ denotes the maximum of $\theta$, so we have $h = \frac{a}{\tan\theta_0}$. When $\alpha_{t}+\alpha_{tr} > 2$, we can differentiate ${A}$ with respect to $\theta_0$, and the zeros of the derivative satisfy 
\begin{equation}
    \cos^{\frac{\alpha_t+\alpha_{tr}-2}{2}}\theta_0\left(\cos^2\theta_0+\frac{\alpha_t+\alpha_{tr}}{2}\sin^2\theta_0\right) = 1.
\end{equation}
Within the range $(0,\frac{\pi}{2})$, the equation has exactly one solution $\hat{\theta}_0$, and this zero corresponds to the maximum of the ideal amplitude gain. Therefore, there exists an optimal distance $h = \frac{a}{\tan \hat{\theta}_0} $ between the feed and the metasurface.
\end{appendices}

\vfill

\end{document}